\documentclass[aps,pra,twocolumn,floatfix,reprint,superscriptaddress, showpacs]{revtex4-1}%,twocolumn,draft
\usepackage{amsmath}
\usepackage{amssymb}
\usepackage[usenames,dvipsnames,svgnames,table]{xcolor}
\usepackage{graphics}
\usepackage{graphicx}
\usepackage{braket}
\usepackage{todonotes}
\usepackage{braket}

\begin{document}

\title{
Full two-photon downconversion of just a single photon}
\author{E. S\'anchez-Burillo}
\affiliation{Instituto de Ciencia de Materiales de Aragon and Departamento de Fisica de la Materia Condensada, CSIC-Universidad de Zaragoza, E-50012 Zaragoza, Spain}

\author{L. Mart\'in-Moreno}
\affiliation{Instituto de Ciencia de Materiales de Aragon and
  Departamento de Fisica de la Materia Condensada, CSIC-Universidad de
  Zaragoza, E-50012 Zaragoza, Spain}

\author{J. J. Garc\'ia-Ripoll}
\affiliation{Instituto de Fisica Fundamental, IFF-CSIC, Calle Serrano
  113b, Madrid E-28006, Spain}

\author{D. Zueco}
\affiliation{Instituto de Ciencia de Materiales de Aragon and Departamento de Fisica de la Materia Condensada, CSIC-Universidad de Zaragoza, E-50012 Zaragoza, Spain}
\affiliation{Fundacion ARAID, Paseo Maria Agustin 36, E-50004 Zaragoza, Spain}

%%%%%%%%%%%%%%%%%%%%%%%%%%%%%%%%%%
%%%%%%%%%%% abs     %%%%%%%%%%%%%%%%%
%%%%%%%%%%%%%%%%%%%%%%%%%%%%%%%%%%

\begin{abstract}
 
  We demonstrate, both numerically and analytically, that it is
  possible to generate two photons from one and only one photon.
  We characterize the output two photon field and make our
  calculations close to reality by including losses.
  Our proposal relies on real or artificial three-level atoms with a
  cyclic transition strongly coupled to a one-dimensional
  waveguide.
  We show that close to perfect downconversion with efficiency over
  $99\%$ is reachable using state-of-the-art Waveguide QED
  architectures such as photonic crystals or superconducting circuits.
In particular, we sketch an implementation in  circuit QED, where
the three level atom is a transmon.
\end{abstract}


\maketitle

%%%%%%%%%%%%%%%%%%%%%%%%%%%%%%%%%%
%%%%%%%%%%% intro
%%%%%%%%%%%%%%%%%%%%%%%%%%%%%%%%%%

\section{Introduction}

The interaction between the electromagnetic field and quantum discrete level systems (like atoms) 
may be enhanced by confining light in one-dimensional waveguides\
\cite {Astafiev2010, Hoi2011,  Hoi2013,  VanLoo2013, Hoi2013b,
  Reitz2013, Thompson2013, Arcari2014, Yalla2014, Goban2014,
  Lodahl2015, Javadi2015}. In these setups, a key parameter is the
ratio between the decay rate due to coupling to waveguide photons and
that due to  coupling to all other channels. Whenever the former dominates, we are in \emph{strong
  coupling} regime of light-matter interactions.
In this case, a single two-level system  can not  be only used to
induce effective photon-photon interactions, but it also enables
\textit{minimal} and \textit{highly efficient} optical devices, such
as perfect mirrors \cite{Fan2005a, Fan2005b,
  Nori2008a}, single photon lasing\ \cite{Fan2012} and Raman
scattering\ \cite{Koshino2013, Inomata2014, Burillo2014}.

Another optical process that could strongly benefit from an enhanced
light-matter interaction is photon downconversion, where a light beam
of a given frequency is split into two beams whose frequencies add up
to the original one. Downconversion is routinely used for the
generation of entangled photons, and light at convenient frequencies. This is already done in atomic and molecular experiments
and it could also be useful for energy harvesting, by using photons of
high energy to excite more suitable transitions in a photovoltaic
material.
Photon down- and up-conversion are currently realized in bulk optics
with the help of nonlinear noncentrosymmetric materials\
\cite{Boyd2003}. Moreover, due to the smallness of the fine structure
constant, the typical performance of this process in crystals such as
BBOs is very small, with only about one in every $10^{12}$ photons being downconverted \cite{Boyd2003}.

A cyclic three level system (C3LS) strongly coupled to a waveguide is
the \textit{minimal} setup that produces downconversion.  When
classical light is used as input, only a small part of the incident
power is converted into a correlated two-photon output field
\cite{Liu2005, Liu2014, Sathyamoorthy2015}.  In chiral waveguides,
however, it has been argued that two photons can be generated when one
and only one photon is scattered in a C3LS structure
\cite{Koshino2009}.  Other downconversion mechanisms at the single
photon limit, requiring the driving of nonlinear cavities, has been
recently proposed \cite{Chang2015}.  In this paper we generalize the
results in C3LS, considering full downconversion efficiency in non
chiral waveguides.  More precisely, we consider a waveguide photon
impinging on the C3LS and resonantly populating level $|2\rangle$, as
schematically represented in Fig. 1. Additionally to the direct
relaxation of $|2\rangle$ to the ground state, the cascade
$|2 \rangle \to | 1 \rangle \to |0\rangle$ allows the relaxation to be
accompanied by the emission of two photons \cite{Scully1997}.
In our study  we
include losses, analyze the entanglement of  the output field  and
suggest  a possible experimental realization.

%%%%%%%%%%%%%%%%%%%%%%%%%%%%%%%%%%%
\begin{figure}[t]
\includegraphics[width=1\linewidth]{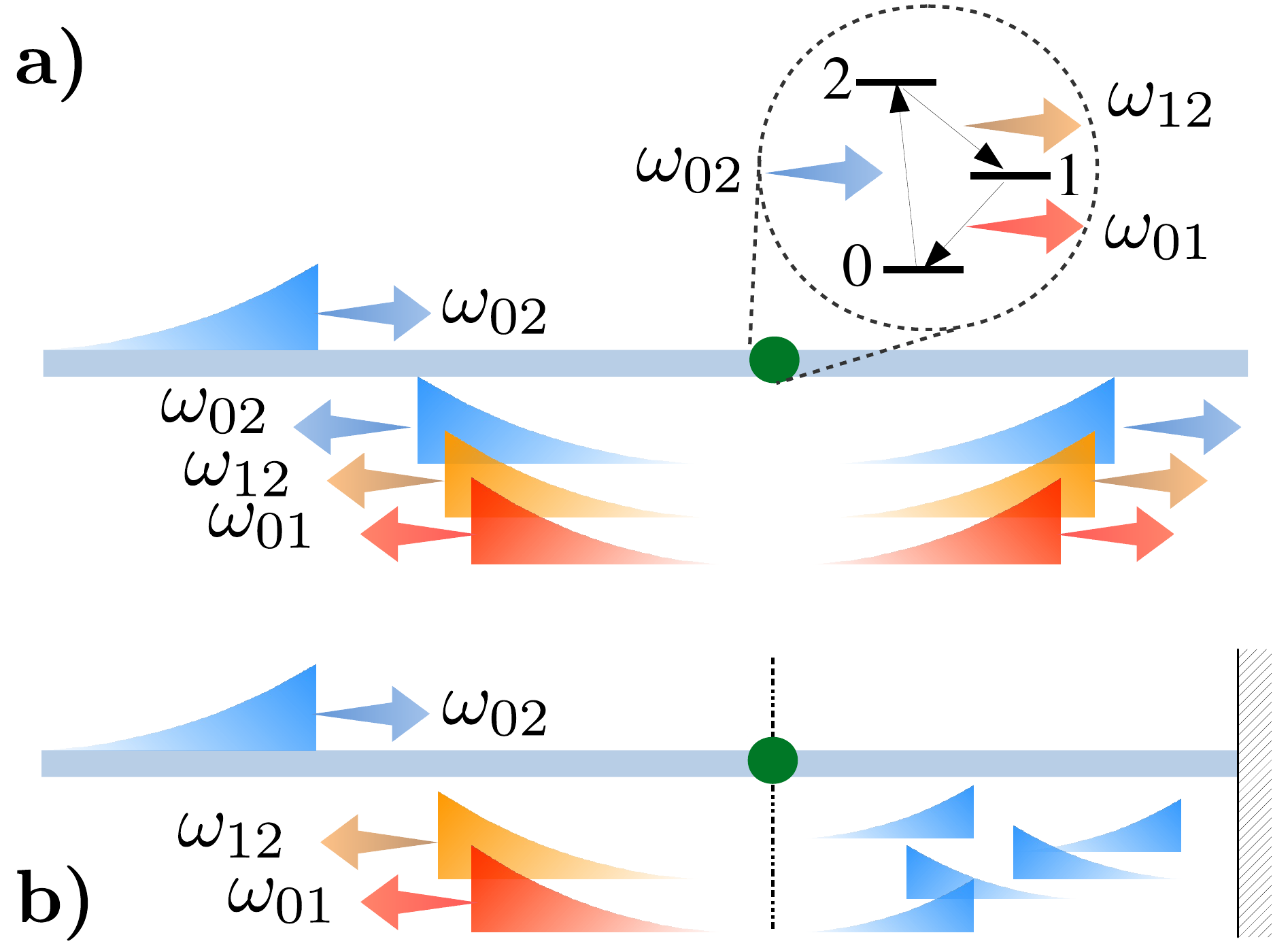}
\caption{(Color online) Downconversion setup. (a) A single incoming
  photon interacts with the three-level system. Part of it is
  transmitted/reflected ($\omega_{02}$, blue) and part is
  downconverted into a pair of photons with frequencies $\omega_{12}$
  and $\omega_{01}$ (orange and red). (b) Placing a mirror right after
  the scatterer at a suitable distance, downconversion can become
  deterministic: all reflected photons have downconverted frequencies. } \label{fig:sketch}
 \end{figure}
%%%%%%%%%%%%%%%%%%%%%%%%%%%%%%%%%%%

The rest of the paper is organized as follows. In the next section, we
introduce the model.  Then, in Sect. \ref{sec:cQED} we sketch a
realization in circuit QED.   We continue by reporting our numerical
results, based on matrix product states (MPS).  There, we discuss the two photon probability
and the dynamics both for the field and the atom. We also characterize
the output field and its entanglement. In Sect. \ref{sec:ana} we
develop an analytical theory, wich allows to compute the efficiency
in presence of losses [Sect. \ref{sec:eff}].  We conclude with the
conclusions and send some technical issues to three appendices.

%%%%%%%%%%%%%%%%%%%%%%%%%%%%%%%%%%
%%%%%%%%%%% model  %%%%%%%%%%%%%%%%%
%%%%%%%%%%%%%%%%%%%%%%%%%%%%%%%%%%

\section{Model}
\label{sec:model}

We consider a cyclic three level quantum system (C3LS) strongly coupled to a one-dimensional waveguide where photons can freely travel. We neglect thermal fluctuations and losses in the waveguide and, for the moment in the C3LS, so the effective Hamiltonian is
($\hbar =1$)
\begin{equation}
H = H_{0} + H_\text{int},
\end{equation}
where
\begin{equation}
\label{H} 
H_{0} =
\int {\rm d}  \omega
\,  \omega   \,  r_\omega^\dagger r_\omega +
\int {\rm d}  \omega  \, \omega  \,   l_\omega^\dagger l_\omega +\sum_{j=0}^2 \omega_j \ket{j}\bra{j},
\end{equation}
with  $r_\omega$ and $l_\omega$ being bosonic operators that, respectively, annihilate right- and left- moving waveguide photons; $r_\omega^\dagger$ and 
$l_\omega^\dagger$ are the corresponding creation operators, and $\omega_j$  and $\ket{j}$ are the eigenenergies and eigenstates of the isolated 3LS.
The coupling between the 3LS and the waveguide photons is represented by $H_\text{int}  = G \; X  $,
with $X$ the electromagnetic (EM) displacement given by
\begin{equation}
X =  \int {\rm d} \omega \, D(\omega)\,  (r_\omega  + l_\omega) +\mathrm{H.c.}
\end{equation}
where $D(\omega)$ is the density of states. 
The operator $G$ accounts for the transitions between levels in the C3LS induced by the EM field:
\begin{equation}
\label{S}
G = g_{01}\ket{0}\bra{1} + g_{12}\ket{1}\bra{2} + g_{02}\ket{0}\bra{2} +\mathrm{H.c.}
\end{equation}

%%%%%%%%%%%%%%%%%%%%%%%%%%%%%%%%%%
%%%%%%%%%%% implementation  %%%%%%%%%%%%%%%%%
%%%%%%%%%%%%%%%%%%%%%%%%%%%%%%%%%%

\section{A possible implementation}
\label{sec:cQED}

An important point is that a C3LS cannot be realized in systems where
(i) quantum states are labelled by a spatial parity tag and (ii) are
small enough and the dipolar interaction dominates (like atoms). The
reason is that at least two of the three states in the C3LS must have
the same parity, but the dipole interaction only couples states with
different parity.  However, effective C3LSs may appear in extended
quantum systems, where couplings beyond the dipolar must be
considered. Implementations of C3LS are some molecules \cite{Kral2001}
and flux qubits made of superconducting circuits \cite{Liu2005,Liu2014}.
However, this last system leads to three quite dissimilar excitation energies.
We chose  an alternative design for
 an effective C3LS in the microwave range using a transmon (a charge
 superconducting qubit shunted by a big capacitor)  that makes the
 C3LS transitions more harmonic \cite{Koch2007}.

%%%%%%%%%%%%%%%%%%%%%%%%%%%%%%%%%%%
\begin{figure}[t]
\includegraphics[width=0.99\columnwidth]{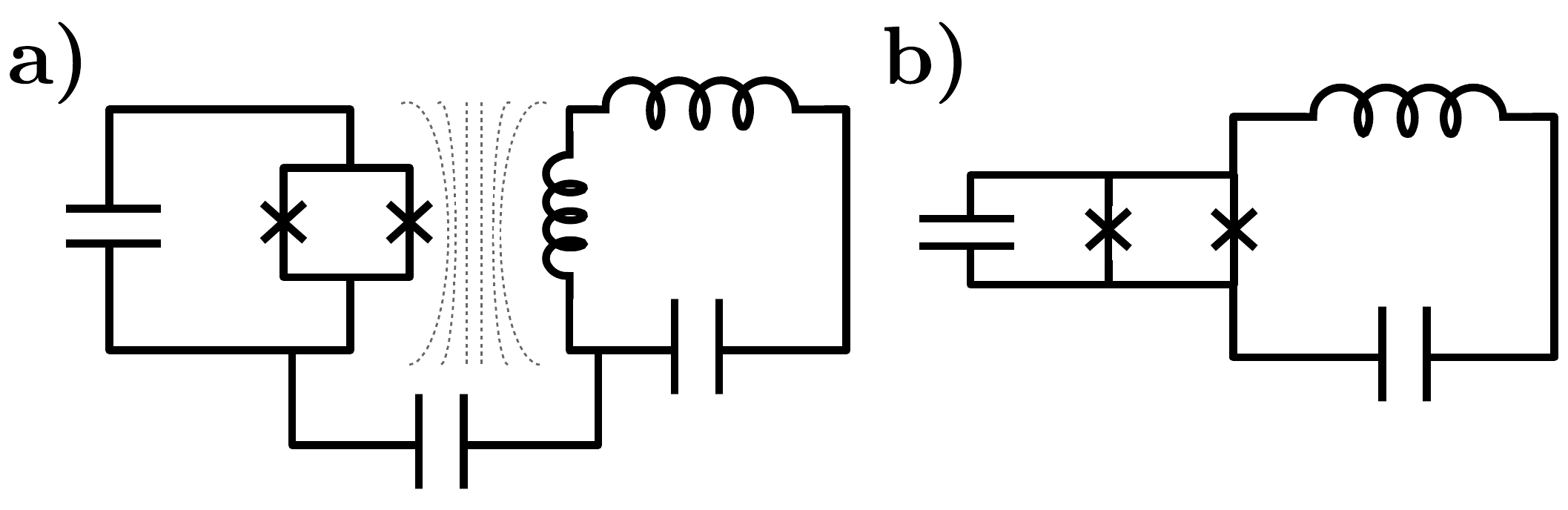}
 \caption{(a) A transmon can be both inductively and capacitively
   coupled to an LC resonator. Coupling strenght can be increased by
   either increasing the SQUID area or (b) by sharing a conductor
   segment, in the spirit of Ref. \cite{Peropadre2013} and similar proposals. }
 \label{fig:circuit}
 \end{figure}
%%%%%%%%%%%%%%%%%%%%%%%%%%%%%%%%%%%

Typically, inductive coupling between the transmon and the
transmission line is negligible. 
The reason is that the transmon design is basically that of a
one-dimensional electric dipole, without support for currents.  In
addition to this, the SQUID that controls the transmon frequency is
small and shielded away from any coupling with the transmission
line. Inductive couplings between tranmons have been however
demonstrated \cite{Chen2014, Dumur2015}.  We make use of similar ideas
to envision a different coupling architecture that allows breaking the
parity symmetry in the transmon setup.

Our starting point is a setup such as the one in
Fig. \ref{fig:circuit}a, where the transmon SQUID is no longer
screened an the superconducting island couples both capacitively and
inductively to the resonator.  
The circuit Lagrangian (with
 inductive  and capacitive  coupling) is,
\begin{align}
\label{l-tr}
{\mathcal L}
=& 
\int {\rm d} x 
\; 
c (\partial_t \phi (x,t) )^2 
-
\frac{1}{l} (\partial_x \phi (x,t) )^2 
\\ \nonumber
& +
\frac{1}{2 C_{\Sigma}} (q -  {\mathcal Q} )^2 - E_J 
\cos (
2 \pi \Phi / \Phi_0 ) \cos \varphi
\, .
\end{align}
The first line accounts for the transmission line Lagrangian.
Here, $\phi(x,t)$ is the (quantum) flux field, that in the interaction
picture reads,
\begin{align}
\phi(x,t) = 
\sqrt{\frac{\hbar Z_0}{4 \pi}}
\int_0^\infty
\, {\rm d} \omega 
\,
\frac{1}{\sqrt{\omega}}
\Big ( 
&  r_\omega {\rm e}^{-i \omega  (t - x/v)}
\\ \nonumber
 & +
 l_\omega {\rm e}^{-i \omega  (t + x/v)}
+
{\rm H.c.}
\Big )
\, ,
\end{align}
with  $c$ ($l$), the capacitance (inductance) per unit length and $Z_0
= \sqrt{l/c}$ is the line  impedance. 
The transmon  and its coupling is written in the  second line.
There,  $E_J$ is the Josephson energy and $C_\Sigma$ is the capacitance.
Charge and phase invariant gauge are quantized via
$[{\rm e}^{i \varphi}, q] = 2 e \; {\rm e}^{i
  \varphi}$.  The transmon 
is  driven and coupled to the line
via the charge  $ {\mathcal Q}$ and the flux  $\Phi$ ( $\Phi_0 = h / 2 e
$ is the flux quantum):
\begin{align}
{\mathcal Q} & = 2 e \, n_g + c \, \partial_t \phi (x,t)
\\
\Phi & = \lambda \,  \partial_x \phi (x,t) + \frac{\Phi_0}{2 \pi}  \varphi_{\rm ext}
\, .
\end{align}
We have introduced the coupling factor $\lambda$ that accounts for the effective field by the transmon's SQUID after taking into account the screening.
Inserting the latter in \eqref{l-tr} and expanding the cosine we get
the coupling Hamiltonian,
\begin{equation}
\label{Hcoup}
H_{\rm coupling}
=
\frac{c}{C_\Sigma}
q \partial_t \phi
-
\lambda d 
\frac{\pi}{\Phi_0} 
E_J
\sin (\varphi_{\rm ext} )
\cos (\varphi) \partial_x \phi.
\end{equation}

%%%%%%%%%%%%%%%%%%%%%%%%%%%%%%%%%%%
\begin{figure}[t]
\includegraphics[width=0.99\columnwidth]{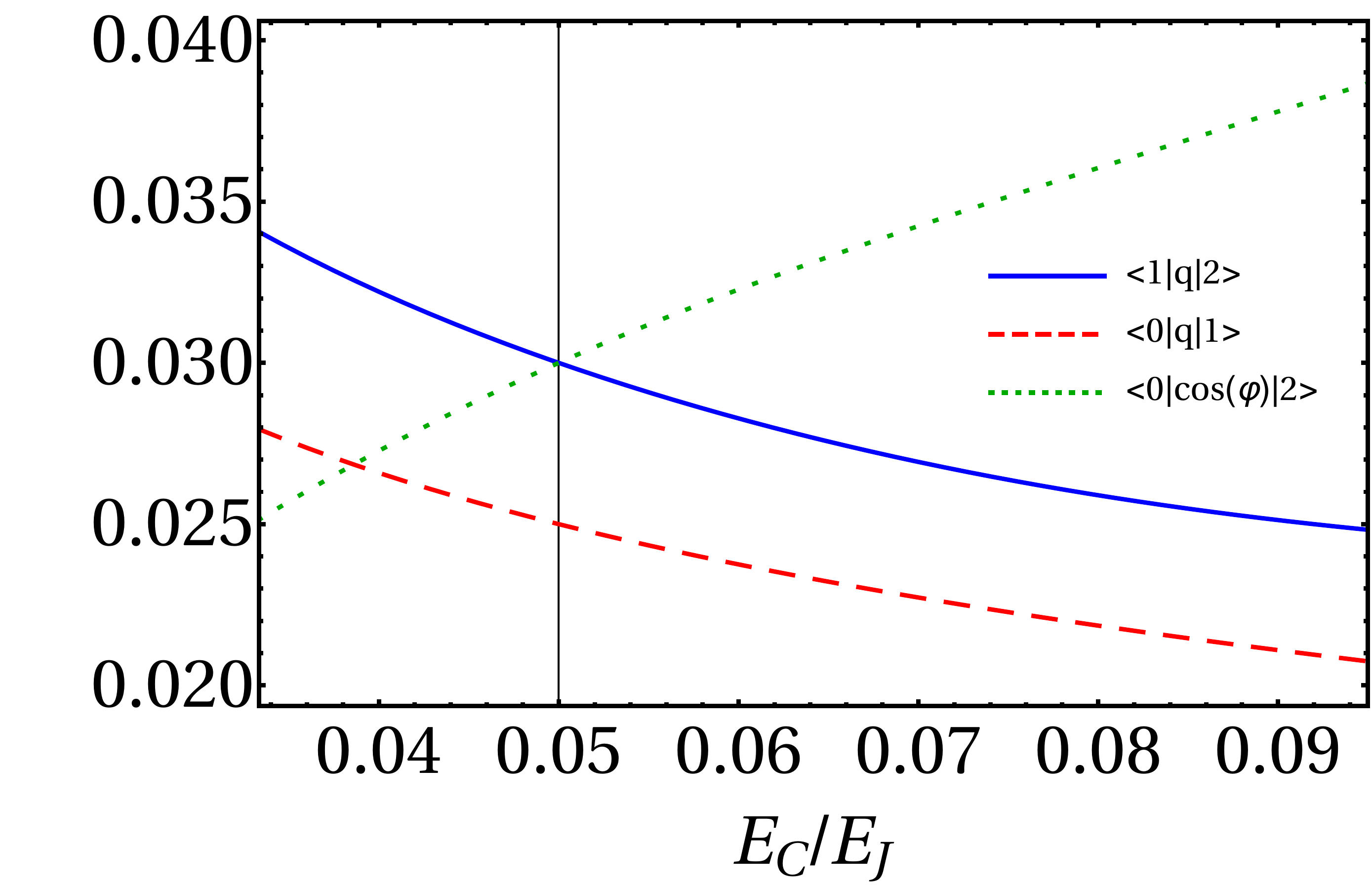}
 \caption{(Color online)  Non zero charge and flux matrix elements $\langle i |
   q | j \rangle$  and $\langle i | \cos \varphi | j \rangle$
   respectively contributing to the coupling operator $G$.  The
   vertical line marks the parameters chosen in our simulations with
   $E_C / E_J=1/20$.  }
 \label{fig:cyclic}
 \end{figure}
%%%%%%%%%%%%%%%%%%%%%%%%%%%%%%%%%%%

We still need to show that \eqref{Hcoup} provides the cyclic structure. 
We numerically diagonalize 
$H_{\rm transmon}= 
\frac{1}{2 C_{\Sigma}} q ^2 - E_J\cos \varphi$
in the charge basis,  retaining the first three levels $H_{\rm
  transmon} = \sum_{j=0}^2 \omega_{j} |j \rangle \langle j |$.  
With the eigenstates at hand,  we can compute the different contributions
to $G$ in $H$ [Eq. (1) in the main text]. 
%
%First, let us comment the part of our results already discussed in the
%literature.
%
In Figure \ref{fig:cyclic}  we plot the contributions
due to  the  charge operator $q$ in \eqref{Hcoup}.  As already
explained in the literature, 
  $\langle i |q | i+1 \rangle \neq 0$ but $\langle 0 | q | 2
 \rangle  = 0$  \cite{Koch2007}. 
The necessary non zero $g_{02}$ value is obtained through the
inductive coupling.   The values for  $\langle 0 | \cos (\varphi)  | 2 \rangle \neq
0$  is also plotted in   \ref{fig:cyclic} ($\langle i | \cos (\varphi) |  i+1 \rangle
= 0$).
Therefore, by combining inductive and capacitive (electric and magnetic) coupling the transmon
has a cyclic structure. 
Through the main text we set $E_C / E_J = 1 /20$.  We fix $\lambda$
and $C_\Sigma$  making  the transition rates between quantum levels induced
by coupling to the waveguide photons, 
 $\Gamma^{(0)}_{ij} \equiv 2 \pi D^2 (\omega_{ij}) g_{ij}^2$,
 optimal for the two photon generation (see below).

%%%%%%%%%%%%%%%%%%%%%%%%%%%%%%%%%%
%%%%%%%%%%% numerics  %%%%%%%%%%%%%%%%%
%%%%%%%%%%%%%%%%%%%%%%%%%%%%%%%%%%
\section{Numerical solution}
We compute the time evolution of an initial single-photon wavepacket.
Even though, the results are independent of the actual wavepacket, our
numerical simulations assume the incident photon being generated via
spontaneous emission in an auxiliary two level system (a single photon
generator).
 Besides, we discretize both space and time and use  the Matrix
 Product States (MPS) technique, which is a well known method for
 obtaining the ground state and low energy states in interacting
 one-dimensional systems
 \cite{Vidal2003,Vidal2004,JJRipoll2006,Verstraete2008,Peropadre2013}.
 MPS  has  been applied to photon scattering in waveguides
 \cite{Burillo2014, Sanchez-Burillo2015, Ramos2016}.
This method is specially suited for Hamiltonians like \eqref{H} that
either have a nonlinear dispersion relation or, as in the considered
case, do not conserve the number of excitations.
It is worth to emphasize here that we solve the time
evolution for the full Hamiltonian. As a consequence, we have access to  both
field and system observables at any time.  Technical details of our
simulations can be found in App. \ref{sec:MPS}.

%%%%%%%%%%%%%%%%%%%%%%%%%%%%%%%%%%%
%%%%%%%%%%%%%%%%%%%%%%%%%%%%%%%%%%%
%%%%%%%%%%%%%%%%%%%%%%%%%%%%%%%%%%%
\subsection{Two photon generation: scattering and dynamics}
%%%%%%%%%%%%%%%%%%%%%%%%%%%%%%%%%%%
\begin{figure}
\includegraphics[width=0.98\columnwidth]{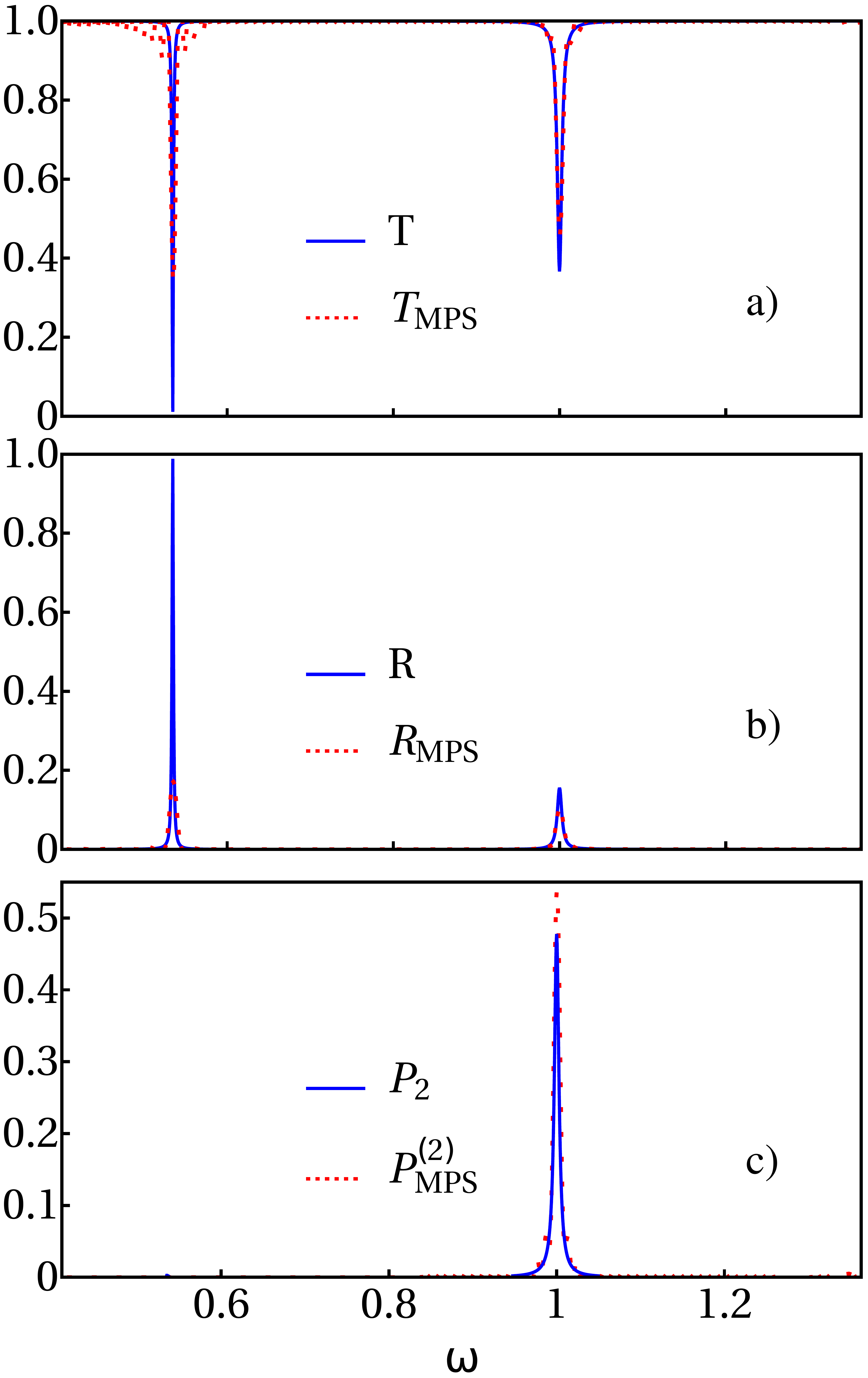}
 \caption{(Color online). Scattering coefficients in a cyclic three-level system, as a function of the incident frequency $\omega$.
One photon transmittance (panel (a)), reflectance (panel(b)) and energy transferred into the two photon channel
$P^{(2)}(\omega)$ (panel(c)).  We
show both analytical (solid lines) and numerical results obtained
with MPS (dotted lines). The parameters are $\omega_{01}=0.59$,
$\omega_{02}=1$, $\Gamma^{(0)}(\omega_{01}) =1.7\times 10^{-3}$,
$\Gamma^{(0)}(\omega_{02})=2.3\times 10^{-3}$ and
$\Gamma^{(0)}(\omega_{12})=3.5\times 10^{-3}$.
We remind that $\Gamma^{(0)}_{ij} = 2 \pi D^2 (\omega_{ij}) g_{ij}^2$.
 }
 \label{fig:T}
 \end{figure}
%%%%%%%%%%%%%%%%%%%%%%%%%%%%%%%%%%%

In Fig. \ref{fig:T} we plot the spectrum for the one photon transmittance and reflectance,
$|t^{(1)}(\omega)|^2 =  \lim_{t\to \infty}|\langle\Omega | r_\omega {\rm
  e}^{-i H t} | \psi_{\rm in} \rangle /\langle \Omega | r_\omega | \psi_{\rm in} \rangle |^2$ and $|r^{(1)}(\omega)|^2 =  \lim_{t\to \infty}|\langle \Omega | l_\omega {\rm
  e}^{-i H t} | \psi_{\rm in} \rangle /\langle\Omega | r_\omega | \psi_{\rm in} \rangle |^2$, respectively,  and the total energy radiated in the two-photon channel $P^{(2)}(\omega)$.
The first transmission dip occurs when the photon energy is centered around  $\omega = \omega_{01}\equiv \omega_1-\omega_0$. In this spectral region the $|0\rangle \to |1\rangle
$ is the  only transition available. Thus, the C3LS behaves as an
effective two-level system and the photon is fully reflected at
resonance \cite{Fan2005a,Fan2005b,Nori2008a}. Consequently,
 $P^{(2)}(\omega)=0$ in this frequency
range [Cf. Fig. \ref{fig:T} c)].
In the second transmission dip, located at 
$\omega = \omega_{02}$, the transmittance presents a finite minimum
value, that is close to 0.5. 
  Figs. \ref{fig:T}  b)-c) shows a remarkable $ 50 \%$  downconversion efficiency of the incoming photon into just two (and only two) outgoing photons, with only a very small amount of light being backreflected.

\begin{figure}[t]
\includegraphics[width=1.0\columnwidth]{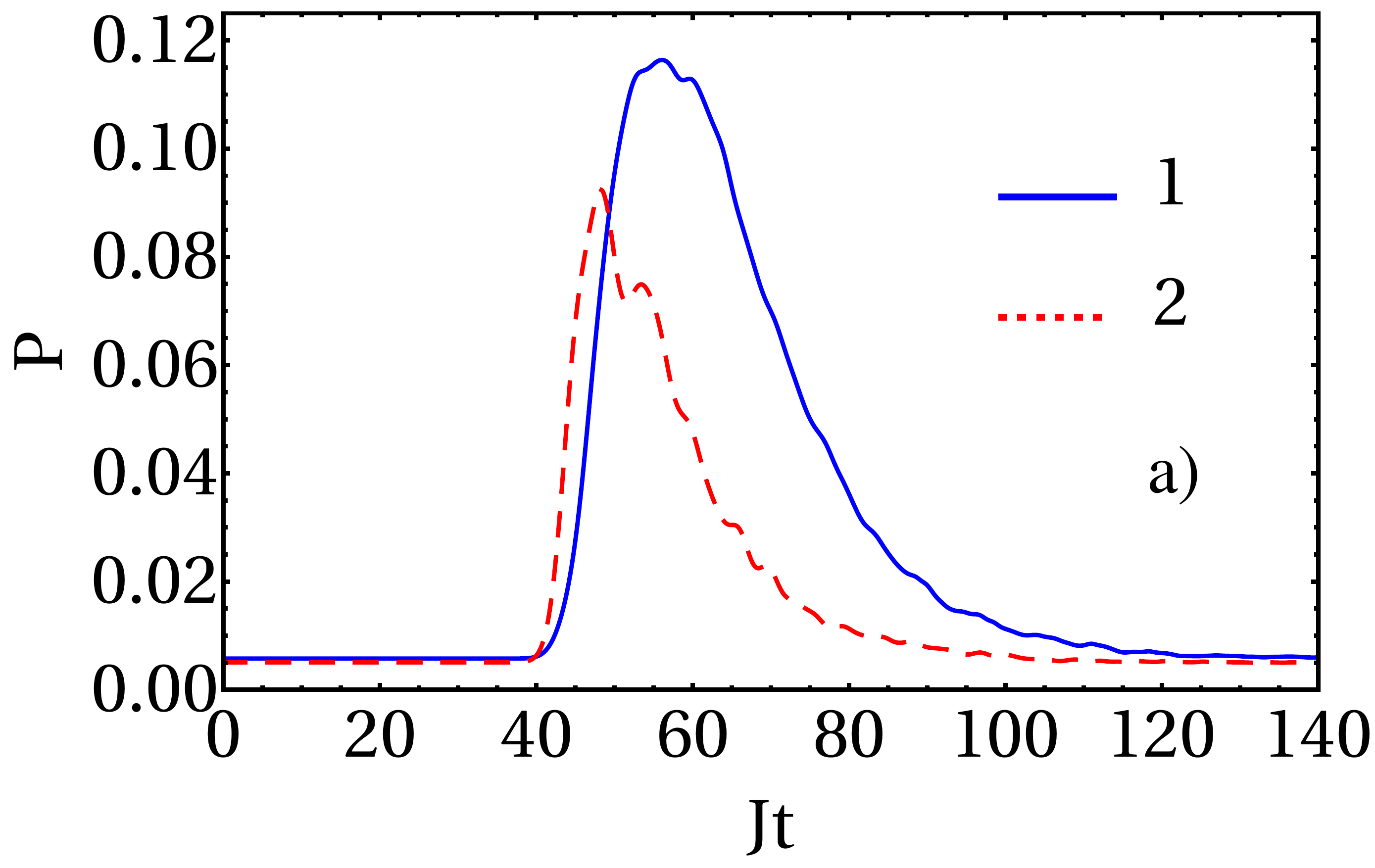}
\includegraphics[width=1.0\columnwidth]{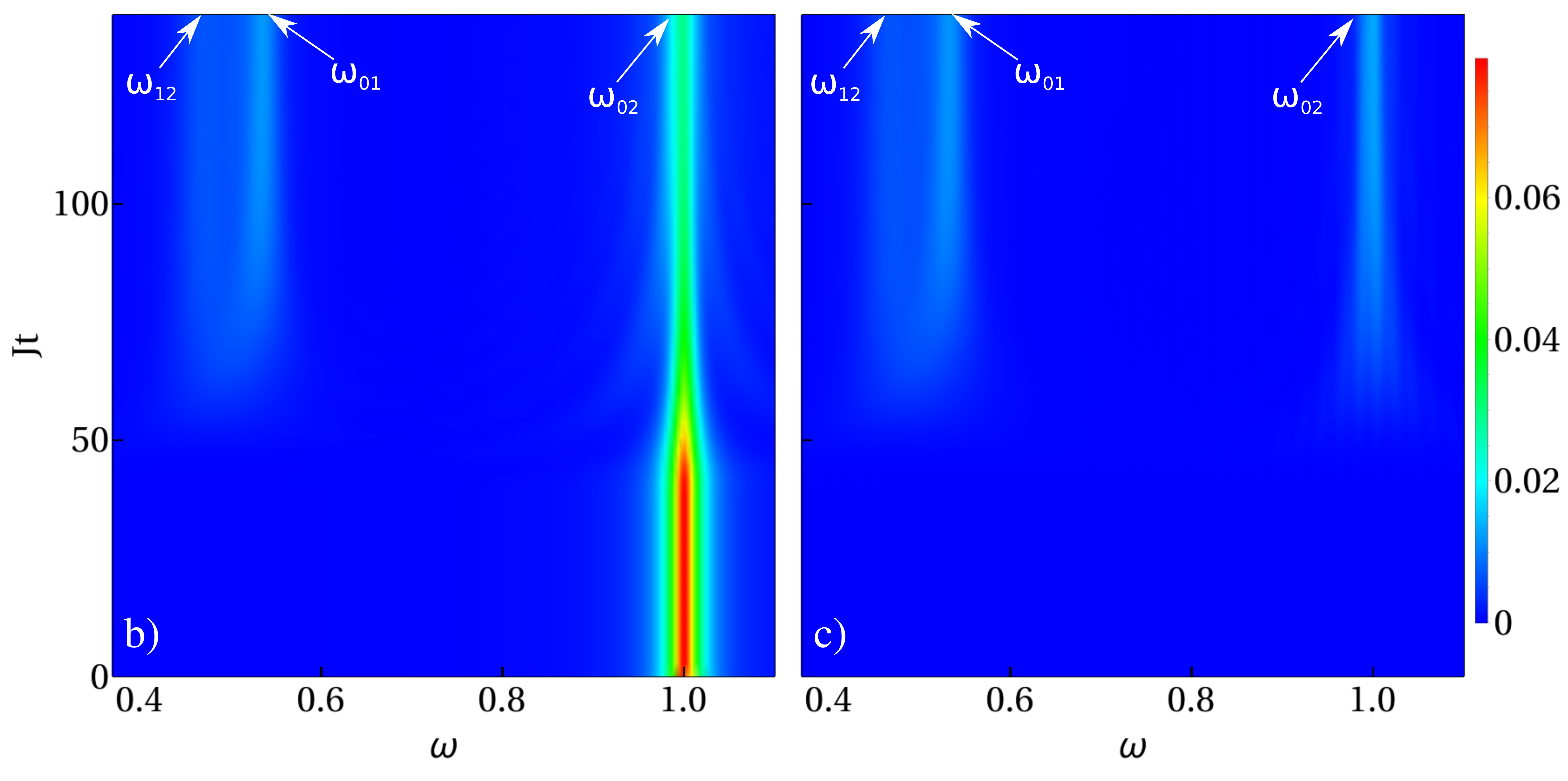}
 \caption{(Color online) a) Population of the first (blue solid line)
   and second (red dashed line) excited states as a function of time. Photon occupation
   in energy space for b) right-moving, $\langle n_\omega^{\text{(r)}}\rangle$, and c) left-moving photons, $\langle n_\omega^{\text{(l)}}\rangle$, respectively, as a function of time. Same parameters as in Fig. \ref{fig:T}.}
 \label{fig:population}
 \end{figure}

For the shake of completeness and to emphasize the fact that we 
have access to the time domain too,  we plot the 3CLS level population in Fig. 
\ref{fig:population} a).  We see that the second excited state gets
populated first,  since our incident photon is resonant
with the transition $\ket{0}\leftrightarrow\ket{2}$. After the
transient  period, both  levels decay to the  ground
state. 
%Notice that these values non-zero, since we are dealing with
%the  full Hamiltonian (we do not use the Rotating Wave approximation)
%. Anyway, their values are  close to zero.
We also plot the particles in energy space, $\langle n^{(r)}_\omega\rangle =
\langle r_\omega^\dagger r_\omega\rangle$ and similarly for  $\langle
n^{(l)}_\omega\rangle $  in
Fig. \ref{fig:population} b) and c). 
In doing so, we can visualize the two photon generation in time domain.
In the beginning, we have a single peak
around the incident energy for the right-moving photons. After the
interaction occurs, a peak appears for a left-moving photon  
at  $\omega_{02}$,  corresponding 
 to the single-photon
reflection [See 
panel \ref{fig:population} c)]. 
In addition, two peaks emerge after the scattering for both 
 forward and backward travelling photons centered at $\omega_{12}$ and
 $\omega_{01}$, associated  to the generation of the two-photon state.

%%%%%%%%%%%%%%%%%%%%%%%%%%%%%%%%%%%
%%%%%%%%%%%%%%%%%%%%%%%%%%%%%%%%%%%
%%%%%%%%%%%%%%%%%%%%%%%%%%%%%%%%%%%

\subsection{Characterization for the two-photon output}

%%%%%%%%%%%%%%%%%%%%%%%%%%%%%%%%%%%
\begin{figure}[t]
\includegraphics[width=0.98\columnwidth]{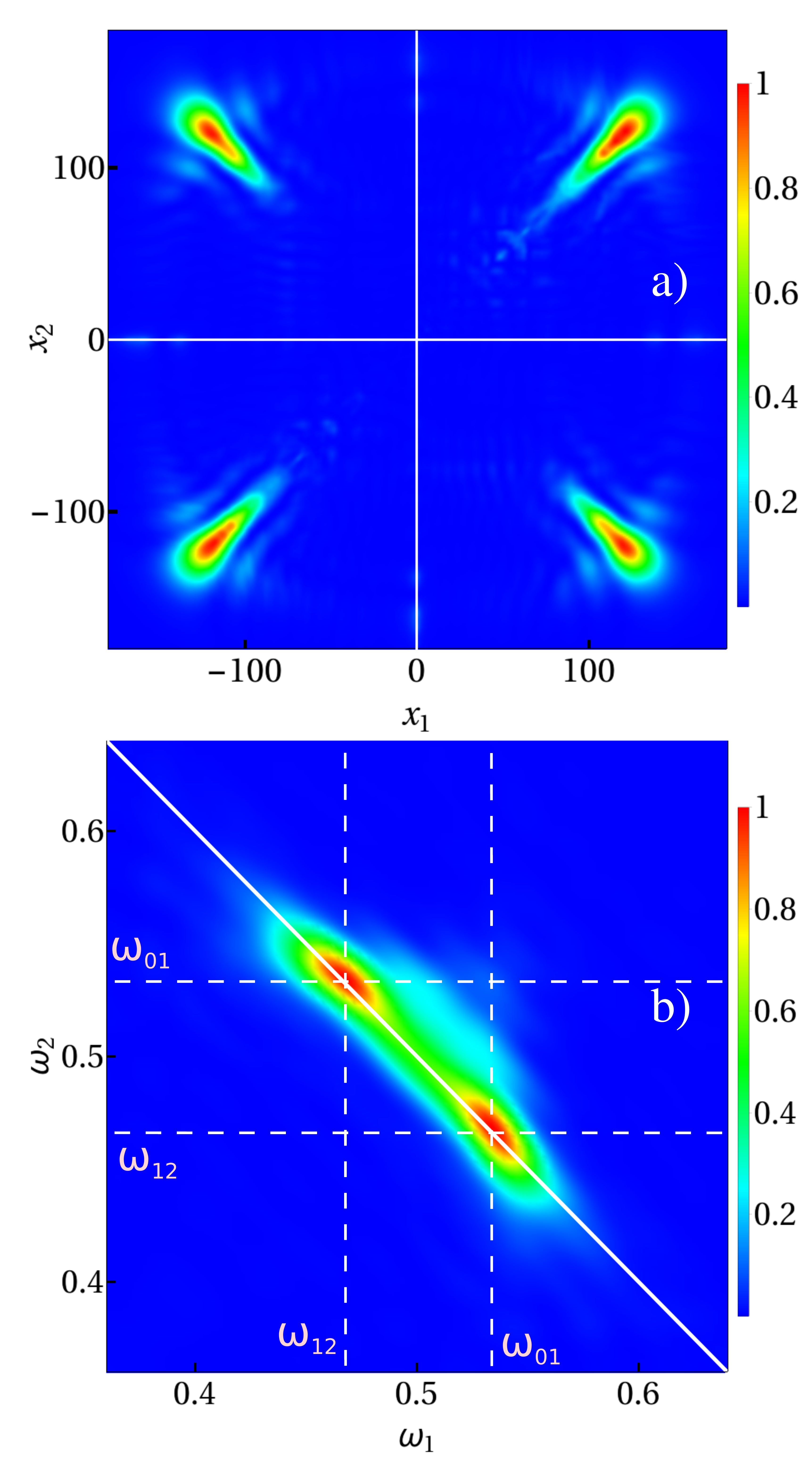}
 \caption{(Color online) Square modulus of the two-photon wave
   function in a) position and b) energy space. The isoenergetic line,
   $\omega_1+\omega_2=\omega$, is shown in the
   bottom panel (white line). We normalize both wave functions such that $\text{max}(|\phi_{x_1x_2}^\text{out}|^2)=\text{max}(|\tilde{\phi}_{\omega_1\omega_2}^\text{out}|^2)=1$. Same parameters as in Fig. \ref{fig:T}.}
 \label{fig:g2}
 \end{figure}

In order to characterize the 
two-photon wave function emerging from the downconversion process we compute the  two-point
correlation function, both in  position space
$\phi_{x_1x_2}^\text{out}:=\langle\Omega
|a_{x_1}a_{x_2}|\Psi(t_\text{out})\rangle$, where $a_x$ 
annihilates a photon at $x$ and in   energy space
for  right-moving photons 
$\tilde{\phi}_{\omega_1\omega_2}^\text{out}:=\langle\Omega |r_{\omega_1}r_{\omega_2}|\Psi(t_\text{out})\rangle$.
As shown in Fig.  \ref{fig:g2}, both photons are emitted spatially in a symmetric way with
respect to the position of the scatterer ($x=0$). In energy space, 
$\tilde{\phi}_{\omega_1\omega_2}^\text{out}$ is centered around
$(\omega_1, \omega_2) =(\omega_{01}, \omega_{12})$ and  $(\omega_{12}, \omega_{01})$ (white dotted lines), as expected from emission from a double resonant process. However, and similarly to the phenomena of resonant fluorescence, $\tilde{\phi}_{\omega_1\omega_2}^\text{out}$ is non-zero all 
along the isoenergetic curve
$\omega_1+\omega_2=\omega$ (white solid line in the  panel b).

\begin{figure}[t]
\includegraphics[width=1.0\columnwidth]{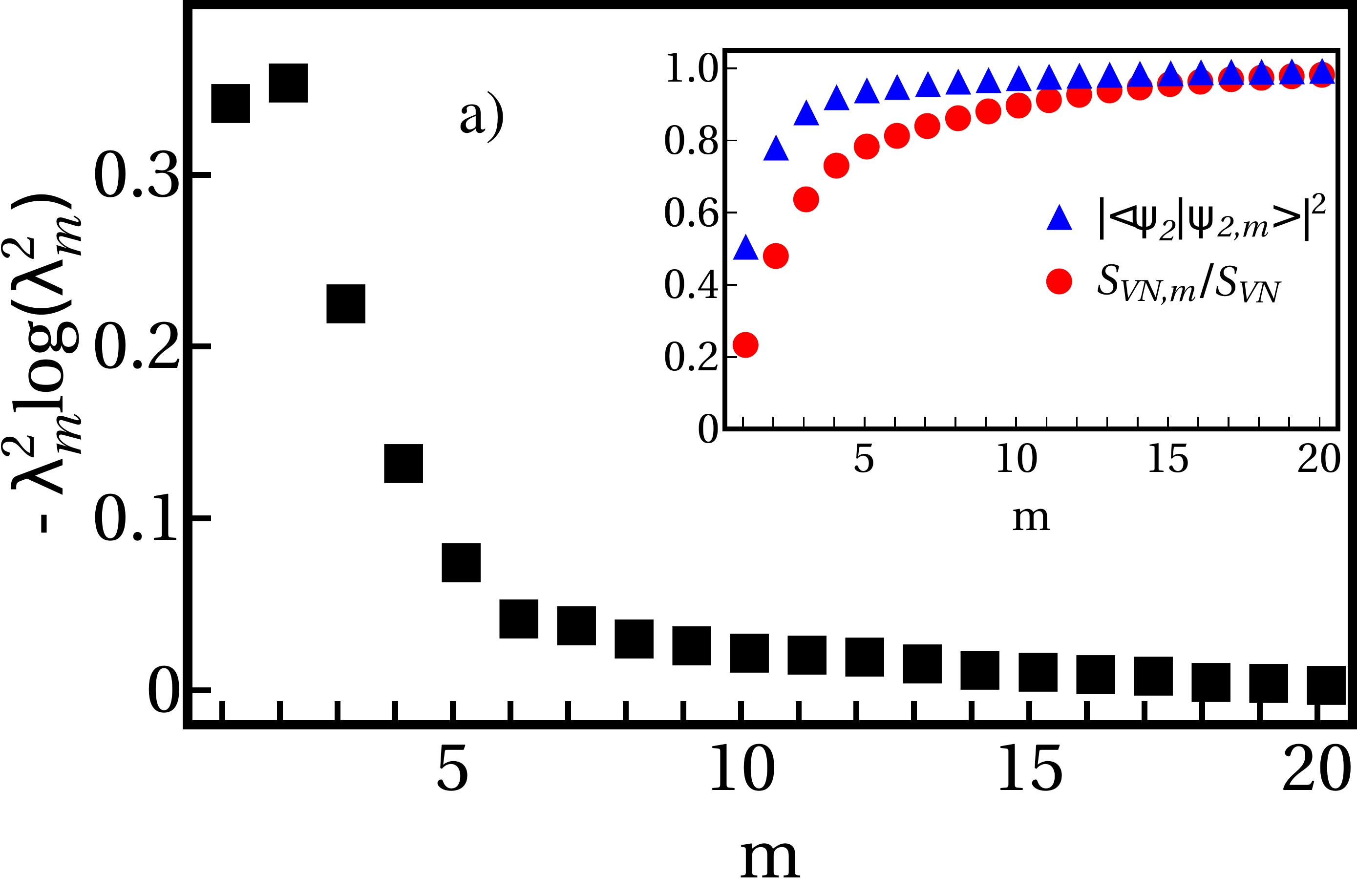}
\includegraphics[width=1.0\columnwidth]{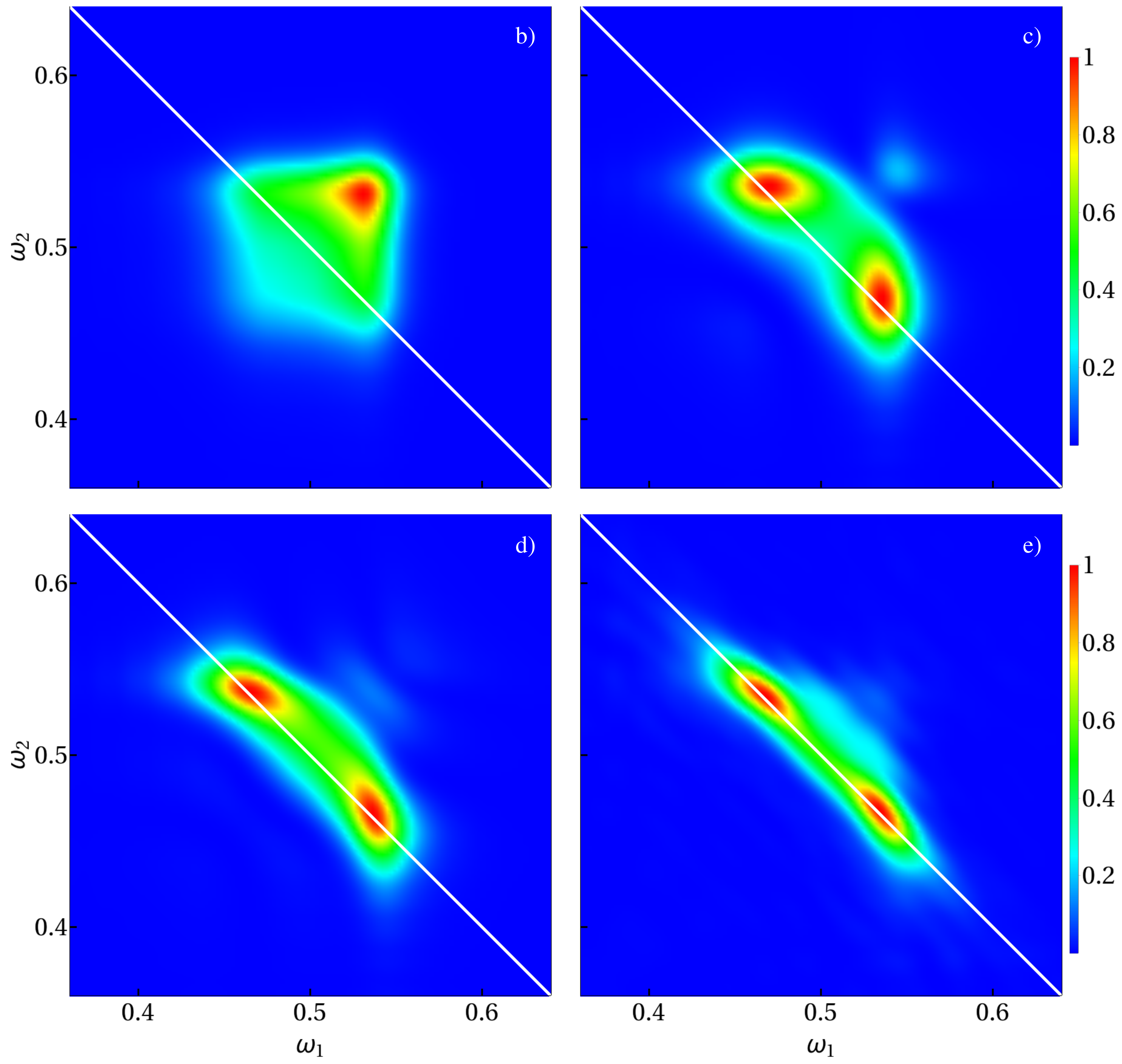}
 \caption{(Color online) a) Contribution of each mode to $S_\text{VN}$,
   $-\lambda_m^2\log(\lambda_m^2)$ as a function of $m$. In the inset,
   we plot the entropy of $|\Psi_{2,m}\rangle$ over the whole entropy,
   $S_{\text{VN},m}/S_\text{VN}$ (red circles) and the overlap between
   $|\Psi_2\rangle$ and $\ket{\Psi_{2,m}}$ (blue triangles) as a
   function of $m$. $|\tilde{\phi}_{\omega_1,\omega_2}^m|^2$ for b)
   $m=1$, c) $m=2$, d) $m=3$ and e) $m=L$  which is the exact result,  Cf. Fig. \ref{fig:g2}b).}
 \label{fig:slogs}
 \end{figure}

The two photons generated are entangled. The corresponding von Neumann entropy $S_\text{VN}$ can be computed after normalizing the two-photon wave function, such that 
 $\sum_{x_1x_2}|\phi_{x_1x_2}^\text{out}|^2=1$, and finding its Schmidt
 decomposition, $\phi_{x_1x_2}^\text{out}=\sum_m \lambda_m
 \varphi_{x_1,m}\chi_{x_2,m}$, being $\{\lambda_m\}$ the singular
 values. Then $S_\text{VN}=-\sum_{m} \lambda_m^2\log(\lambda_m^2)$
\cite{You2001}
. In
 the representative case shown in Fig.  \ref{fig:g2} we get
 $S_\text{VN}=1.44$.   
For a better understanding,  we plot 
the contribution of each mode to $S_{\rm VN}$ in Fig. \ref{fig:slogs}
a). 
The entropy is dominated by the first two modes, but the contribution
from the other modes is non negligible. 
In order to quantify how the entropy is recovered from a given number
of modes, 
 we define the entanglement entropy of the first $m$ modes
\begin{equation}
S_{\text{VN},m}=-\sum_{n=1}^m \lambda_n^2 \log(\lambda_n^2) \, ,
\end{equation}
and  show $S_{\text{VN},m}/S_\text{VN}$ in the inset of
Fig. \ref{fig:slogs}. 
%It converges to $1$ for $m\simeq 20$. 

Another measure of how the wavefunction can be represented by a fixed
number of modes is the fidelity, \emph{i.e.} 
 the overlap between the actual two photon state,
 $|\Psi_2\rangle=1/\sqrt{2}\sum_{x_1,x_2}\phi_{x_1,x_2}a_{x_1}^\dagger
 a_{x_2}^\dagger|\Omega\rangle$ and the state reconstructed with $m$ modes:
\begin{equation}
\label{psim}
|\Psi_{2,m}\rangle=\frac{1}{\sqrt{2}}\sum_{x_1,x_2} \sum_{n=1}^m \lambda_n\tilde{\varphi}_{x_1,n} \tilde{\chi}_{x_2,n}a_{x_1}^\dagger a_{x_2}^\dagger|\Omega\rangle.
\end{equation}
In Fig. \ref{fig:slogs} a) (inset) we check that the overlap
qualitatively behaves as $S_\text{VN,m}$.

Lastly, we can visualize how the  two-point correlation function is
reconstructed by adding modes.  In Fig. \ref{fig:slogs}  we plot  
%two-point correlation function  in frequency space for the state
%\eqref{psim}, namely 
%the two-photon state with a partial reconstruction considering just $m$ modes.
%In terms of the Schmidt decomposition, the two-photon state is
$| \tilde{\phi}^\text{m}_{\omega_1,\omega_2} |^2 = |\langle\Omega
|r_{\omega_1}r_{\omega_2}|\Psi_{2,m}\rangle|^2$ for different values
of $m$.
%$
%%=\sum_{m} \lambda_m
%%\tilde{\varphi}_{\omega_1,m}\tilde{\chi}_{\omega_2,m}$,  with $m=1,\dots, L$.
%We plot $|\tilde{\phi}_{k_1,k_2}^{\text{out},m}|^2$, with
%$\tilde{\phi}_{k_1,k_2}^{\text{out},m}:=\sum_{n=1}^m \lambda_n
%\tilde{\varphi}_{k_1,n}\tilde{\chi}_{k_2,n}$, 
% for b) $m=1$, c) $m=2$,
%d) $m=3$ and e) $m=L$ .  
 The white lines, as in Fig. \ref{fig:g2}, mark the isoenergertic condition. For $m=1$ we do not catch the bimodal
aspect of the state. However, already  with $m=2$ we see the double-peaked structure.

%%%%%%%%%%%%%%%%%%%%%%%%%%%%%%%%%%%%%%%%
%%%%%%%%%%%% Analytical estimation
%%%%%%%%%%%%%%%%%%%%%%%%%%%%%%%%%%%%%%%%
\section{Analytical theory}
\label{sec:ana}  

In order to provide an approximate analytical theory for the numerical results presented above, we
use the input-output formalism \cite{Gardiner1985} that has been
recently introduced to scattering problems in Waveguide QED
\cite{Fan2010, Xu2015}.
The central object in this theory is the relation among the input and
output fields, 
 namely,
\begin{equation}
\label{io}
r_ {\rm out} (t) = r_{\rm in} (t) - i  \sqrt{2 \pi } D(\omega) G(t)
\, ,
\end{equation}
 defined as 
$
r_{\rm in} (t)  :=   \int_0^\infty  \frac {{\rm d} \omega }{\sqrt{2
    \pi}}
\;  
r_{\omega} (t_0) {\rm e}^{-i \omega t}
$
and
$
r_{ \rm out} (t)  := \int_0^\infty  \frac {{\rm d} \omega }{\sqrt{2 \pi}}
r_{ \omega} (t_f) {\rm e}^{-i \omega (t-t_f)}
$ with 
 $r_{\omega} (t) = {\rm e}^{i Ht} r_\omega {\rm e}^{-i Ht} $.
The times $t_0$ and $t_f$ must be taken well
before and after the scattering event has occurred. As we are
interested in  asymptotic behavior,  we can set $t_0 \to -\infty$ and
$t_f \to \infty$.  It is important to notice that, in general, the relation
\eqref{io}  is obtained assuming that the atom spontaneous emission
rates $\Gamma^{(0)}_{ij} = 2 \pi D^2 (\omega_{ij}) g_{ij}^2$ are 
small compared to the bare atom transitions.   

The crucial point, as expressed in Eq. \eqref{io}, is that the full
photon dynamics can be obtained from that of the quantum scatterer by solving the 
quantum optical master equation \cite{Gardiner1985} for the reduced density matrix of the C3LS:
\begin{align}
\label{qme-0}
\frac {d \varrho}{dt}
= &
-i [H_{\rm 0}, \varrho]
- 2 \imath \alpha D(\omega) \cos (\omega t) 
[G(t), \varrho]
\\ \nonumber
&+
2  
\sum_{\omega_{ij} > 0}
\Gamma_{ij}
\Big ( 
L_{ij} \varrho L_{ij}^\dagger 
-
\frac{1}{2}
\{  L_{ij}^\dagger   L_{ij} , \varrho \} 
\Big )  \, ,
\end{align}
where 
$
L _{ij} = |j \rangle \langle i |
$ and $\Gamma_{ij} $ are the transition rates between the discrete levels in the scatterer. Additionally to the transition rates induced by coupling to the waveguide photons 
this formulation allows us to consider the transitions $\gamma_{ij}$ induced by coupling to other baths (as phonons or other components of the EM field). In this case, the total transition rate is $\Gamma_{ij} = \Gamma^{(0)}_{ij} + \gamma_{ij}.$

As stated above, we have numerically tested that no more than two
photons are generated in the dynamics.  Therefore, the two photon generation
probability can be computed  by energy conservation:
\begin{equation}
\label{tau}
P^{(2)}(\omega) =  1- |t^{(1)}(\omega)|^2 -|r^{(1)}(\omega)|^2  -
A(\omega) \, .
\end{equation}
The one photon transmittance  $t_1(\omega)$ is given by
\begin{equation}
\label{t1-c}
t^{(1)} (\omega) = 
\lim_{\alpha \to 0}
\frac
{
 \langle \alpha_\omega | r_{\rm out}  (t) | \alpha_\omega
\rangle
}
{
 \langle \alpha_\omega | r_{\rm in}  (t) | \alpha_\omega
\rangle
}
\end{equation}
with $
| \alpha_\omega \rangle = {\rm e}^{\alpha r_{\rm in}^\dagger(\omega) - \text{H.c.}} | \Omega \rangle$, being $r_{\rm in}^\dagger(\omega)$ the Fourier transform of $r_{\rm in}^\dagger(t)$.  Thus, $t^{(1)} (\omega)$ can be obtained
by solving \eqref{qme-0} within linear response theory.   After some algebra, 
we get \ref{app:p2}

%\begin{widetext}
\begin{equation}
\label{tcyclic}
t^{(1)} (\omega) = 1 -
\frac{ i  \Gamma^{(0)}_{01}} {(\omega - \omega_{01}) + i \Gamma_{01}}
- \frac{ i  \Gamma^{(0)}_{02}}{(\omega - \omega_{02}) + i (\Gamma_{01}+\Gamma_{02})}.
\end{equation}
%\end{widetext}
In addition, $ 1+r^{(1)} (\omega) =  t^{(1)} (\omega) $.

The last term in \eqref{tau} is the energy ``absorbed" by the lossy
channels, $A(\omega)$.  Here, we consider a unique channel dissipating
all the C3LS transitions.
Around the two photon frequency generation $A(\omega)$ can be approximated,
\begin{equation}
\label{tau}
A (\omega \cong \omega_{02}) =  2 \, \gamma_{02} \, 
|r^{(1)} (\omega_{02})|^2 / \Gamma_{02}
\end{equation}
The validity of these approximate analytical expressions is shown in  
Fig. \ref{fig:T}, where they are compared to the numerical results for the ``lossless" case $\gamma_{ij}=0$.

%%%%%%%%%%%%%%%%%%%%%%%%%%%%%%%%%%%%%%%%%%%%
%%%%%%%%%%% efficiency                          %%%%%%%%%%%%%%
%%%%%%%%%%%%%%%%%%%%%%%%%%%%%%%%%%%%%%%%%%%%

\section {Efficiency}
\label{sec:eff}

Equation \eqref{tau} allows for the search of optimal parameters for downconversion.  
The first observation is that losses are detrimental,  always
reducing $P^{(2)} (\omega)$.
Even in absence of losses ($\gamma_{ij}=0$), the two-photon generation can be considered as a loss mechanism {\it for the one-photon channel}, which implies that  the fraction of energy downconverted is at most $\max P^{(2)} (\omega)= \frac{1}{2}$ (occurring when $r (\omega) = -\frac {1}{2}$). 
This fundamental bound is related to the fact that a deep subwavelength scatterer re-emits equally to the left and to the right \cite{Peropadre2011}.
But this bound can be exceeded by breaking the left-right symmetry in the waveguide by, \emph{e.g.} placing a mirror next to the C3LS,
as sketched in Fig. \ref{fig:sketch}b).

The reflectance and (for $\gamma_{ij}\ne 0$) absorption can be calculated in this configuration 
by summing all multiple-scattering processes that the waveguide photon has with both the C3LS and the mirror \footnote {Eventually,
we will set $r_M=-1$, \emph{i.e.} we neglect losses in the mirror, wich is a good
experimental assumption.}.
The sum  can be done
analytically \ref{app:eff}, resulting in:
\begin{align}
\label{mirror}
P^{(2)} (\omega) = 
1
 & - \left | 
\frac{r^{(1)}(\omega) - \Big (1+ 2 r^{(1)}(\omega) \Big ) \Phi(\omega)}{1 + r^{(1)}
  (\omega) \Phi(\omega)}
\right |^2
\\ \nonumber
& -
\left |
\frac{1 - \Phi(\omega)}{1 + r^{(1)}(\omega) \Phi(\omega)} \right |^2 \, A(\omega),
\end{align}
where $\Phi(\omega) =  {\rm e}^{2 i k(\omega) d}$,  
$d$ is the distance between the mirror and the C3LS and $k(\omega)$ is
the waveguide photon wavevector at frequency $\omega$.

%%%%%%%%%%%%%%%%%%%%%%%%%%%%%%%%%%%
\begin{figure}
\includegraphics[width=0.98\columnwidth]{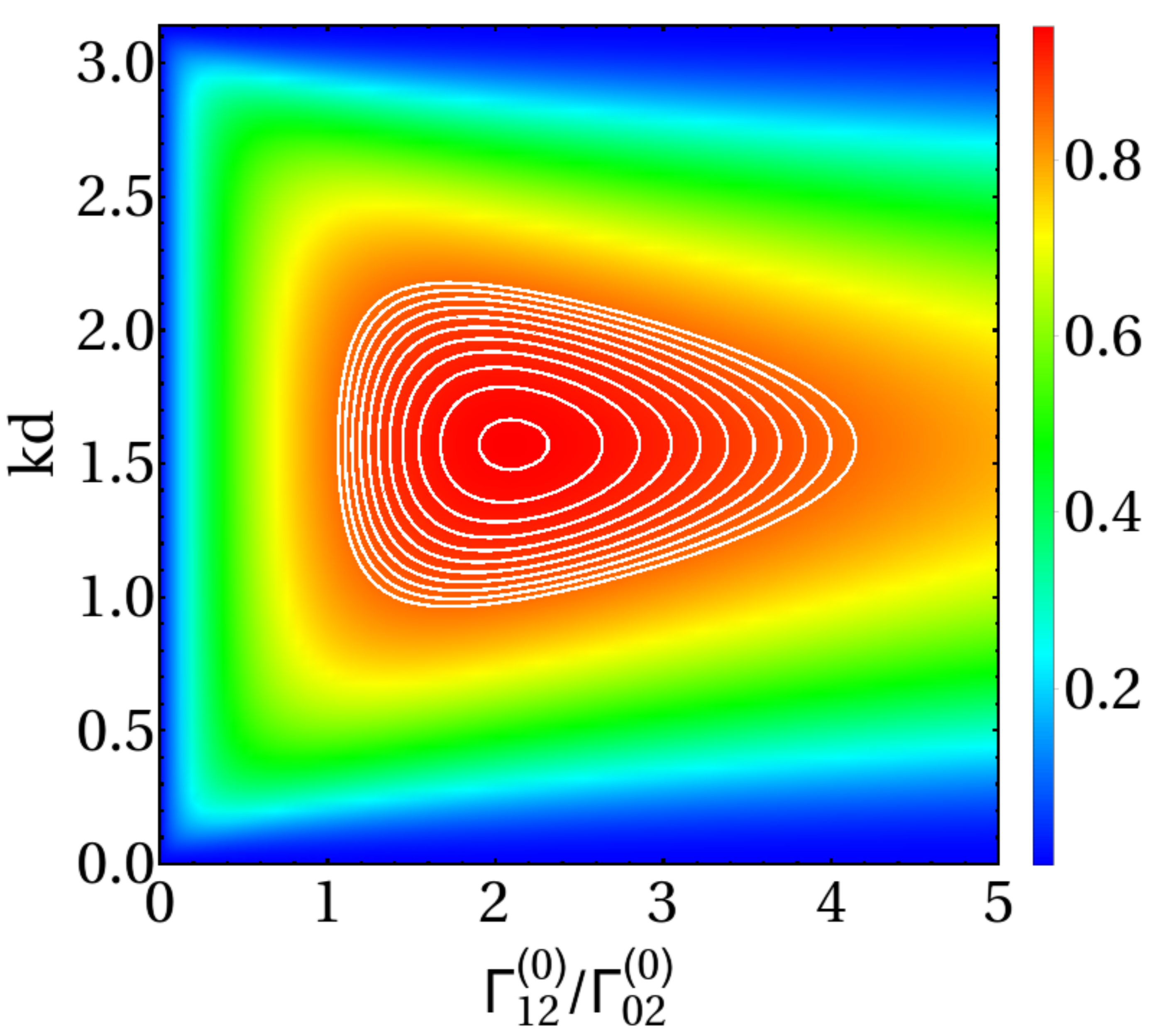}
 \caption{(Color online) $P^{(2)}(\omega = \omega_{02})$ as a function of
   the distance atom-mirror $kd$, See Fig. \ref{fig:sketch}, and the
   ratio $\Gamma^{(0)}_{12} / \Gamma^{(0)}_{02}$.  Losses are taken
   into account.  In the figure, a conservative ratio 
  $\gamma_{02} / \Gamma_{02}^{(0)} = 0.1$ is used.  The rest of the
 parameters are the same as in Fig. \ref{fig:T}.  White lines mark
 iso-efficiency curves, starting at 0.9 and finishing at 0.99.} 
 \label{fig:P2}
 \end{figure}
%%%%%%%%%%%%%%%%%%%%%%%%%%%%%%%%%%%
As drawn in Fig. \ref{fig:P2},  the maximum downconversion efficiency
predicted by Eq. \eqref{mirror} occurs at resonance
($\omega=\omega_{02}$), and for $\Gamma_{12}^{(0)}/\Gamma_{02}^{(0)}
\cong 2 $ and  $k d= \pi/2$, and can be approximated by:
\begin{equation}
\label{maxP2}
{\rm max} P_{2} = 1 - \frac{\gamma_{02}}{ \Gamma_{02}}
\end{equation}

So, remarkably, downconversion may be perfect in the considered configuration if losses are negligible.
It provides a simple expression for the maximum efficiency as a
function of the ratio between the rates for absorption and coupling into waveguide photons.
This ratio is a key  figure of merit in Waveguide QED and values as small as  
 $10^{-2}$ have already reported for effective two-level systems in both superconducting circuits \cite{Arcari2014} and
photonic crystals \cite{Sollner2015}.
Thus, two photon generation with one and only one photon with an efficiency
larger than $0.99$ is doable using an appropriate C3LS.
%
%%%%%%%%%%%%%%%%%%%%%%%%%%%%%%%%%%
%%%%%%%%%%% conclusions     %%%%%%%%%%%%
%%%%%%%%%%%%%%%%%%%%%%%%%%%%%%%%%%

\section{Conclusions}
We have shown that two photons can be efficiently generated by sending one and
only one photon through a  cyclic three-level atom in a realistic scenario.
Remarkably, the downconversion process can occur with unit
probability, being only limited by energy leakage in the three-level
system. 
Based on reported experimental data, we have estimated that a nearly perfect two photon
generator operating at the single photon level is feasible in
architectures based on either photonic crystals or superconducting circuits.
Together with single atomic mirrors \cite{Fan2005a,Fan2005b,Nori2008a}, single photon
lasing \cite{Fan2012}  or single photon Raman scattering 
\cite{Burillo2014}, this work contributes to the toolbox of photonics with minimum power, where even tasks usually associated to high intensities are performed at the one-photon level.

%%%%%%%%%%%%%%%%%%%%%%%%%%%%%%%%%%
%%%%%%%%%%% Acknowledgements    %%%%%%%
%%%%%%%%%%%%%%%%%%%%%%%%%%%%%%%%%%

\begin{acknowledgements}
We acknowledge 
support by the Spanish Ministerio de Economia y Competitividad within
projects MAT2014-53432-C5-1-R, FIS2012-33022 and  FIS2014-55867-P, CAM Research Network QUITEMAD+ and  the Gobierno
de Arag\'on (FENOL group).

\end{acknowledgements}

\begin{appendix}

\section{One photon scattering, input-output and Linear Response Theory}
\label{app:p2}

To start with, we define the $S$-matrix,  as $S := \lim_{t \to \infty}  U(t, -t) $ with $U(t, t^\prime) = {\rm
  e}^{- i H (t -t^\prime)}$ the
evolution operator.
Then \cite{Fan2010, Xu2015}:
\begin{align}
\label{t1-0}
t^{(1)}(\omega) =
\lim_{t\to \infty}
\frac{\langle \Omega  | \, r_{\omega}  r^\dagger_{\rm out}(t)  \,  |\psi_{\rm in}
  \rangle
}
{
\langle \Omega  | \,  r_{\omega}  r^\dagger_{\rm in}(t)  \,   | \psi_{\rm in}
  \rangle 
}
\, .
%\frac{ \langle r_{\rm out} \rangle} {\langle r_{\rm in} \rangle}
\end{align}
The second equality holds after direct replacement of the definitions
for the input output fieds  appearing in the Gardiner and Collet seminal
paper \cite{Gardiner1985}, 
\begin{align}
\label{rin}
r_{\rm in} (t)  &:=   \int_0^\infty  \frac {{\rm d} \omega }{\sqrt{2
    \pi}}
\;  
r_{\omega} (t_0) {\rm e}^{-i \omega (t-t_0)}
\\ \label{rout}
r_{ \rm out} (t)  & := \int_0^\infty  \frac {{\rm d} \omega }{\sqrt{2 \pi}}
r_{ \omega} (t_f) {\rm e}^{-i \omega (t-t_f)}
\end{align}
Here, 
$r_{\omega} (t) = {\rm e}^{i Ht} r_\omega {\rm e}^{-i Ht} $ are
Heisenberg evolved operators.
The times $t_0$ and $t_f$ are times well
before and well after the scatterer and the impinged photons have
interacted.   If we are interested in  asymptotics,  we can set
$t_0 \to -\infty$ and $t_f \to \infty$.

Consider now a coherent input state, 
\begin{equation}
\label{alp-in}
| \alpha_\omega \rangle = {\rm e}^{\alpha r^\dagger_{\rm in}(\omega) - \text{H.c.}} | \Omega \rangle.
\end{equation}
We consider the following expected value
\begin{equation}
f_{\rm out}(\omega,\omega',\alpha):=\braket{ \alpha_\omega | r_{\rm out}  (\omega') | \alpha_\omega},
\end{equation}
with $r_{\rm out}  (\omega)$ the Fourier transform of $r_{\rm out}  (t)$, Eq. \eqref{rout}. We take a series expansion in $\alpha$
\begin{equation}
f_{\rm out}(\omega,\omega',\alpha)=\alpha\braket{\Omega|r_{\rm out}(\omega')r_{\rm in}^\dagger(\omega)|\Omega}+\mathcal{O}(\alpha^2)
\end{equation}
Following Fan et al. \cite{Fan2010}, $\braket{\Omega|r_{\rm out}(\omega')r_{\rm in}^\dagger(\omega)|\Omega}=t^{(1)}(\omega) \delta(\omega-\omega')$. Thus
\begin{equation}
f_{\rm out}(\omega,\omega',\alpha)=\alpha\; t^{(1)}(\omega) \delta(\omega-\omega') +\mathcal{O}(\alpha^2).
\end{equation}
Fourier transforming with respect to $\omega'$
\begin{align}
f_{\rm out}(\omega,t,\alpha):=&\frac{1}{\sqrt {2\pi}} \int \mathrm{d} \omega' f_{\rm out}(\omega,\omega',\alpha) e^{i\omega' t}\\
=&\frac{\alpha}{\sqrt{2\pi}}\; t^{(1)}(\omega)e^{i\omega t}+\mathcal{O}(\alpha^2).
\end{align}
Notice that $f_{\rm out}(\omega,t,\alpha)=\braket{\alpha_\omega|r_{\rm out}(t)|\alpha_\omega}$. Then, the transmission amplitude can be computed as
\begin{equation}
t^{(1)}(\omega)=\lim_{\alpha\to 0}\frac{\braket{\alpha_\omega|r_{\rm out}(t)|\alpha_\omega}}{\braket{\alpha_\omega|r_{\rm in}(t)|\alpha_\omega}}
\end{equation}
%\begin{equation}
%\lim_{\alpha \to 0} \langle \alpha_\omega | r_{\rm out}^\dagger  (t) | \alpha_\omega
%\rangle =
%\langle \alpha_\omega | r_{\rm out}^\dagger  (t) | \Omega \rangle
%\; , 
%\end{equation}
%and  $r_\omega |\alpha_\omega\rangle =
%\alpha_\omega|\alpha_\omega\rangle $ we can re-write \eqref{t1-0},
%\begin{equation}
%\label{t1-c}
%t^{(1)}(\omega) = 
%\lim_{t \to \infty}
%\lim_{\alpha \to 0}
%\frac
%{
% \langle \alpha_\omega | r_{\rm out}^\dagger  (t) | \alpha_\omega
%\rangle
%}
%{
% \langle \alpha_\omega | r_{\rm in}^\dagger  (t) | \alpha_\omega
%\rangle
%}
%\end{equation}
Therefore, the one photon scattering can be obtained by driving the
scatterer with a coherent (classical)  state in the limit of weak
amplitude, $\alpha$.  Thus, Linear Response Theory can be used.

%%%%%%%%%%%%%%%%%%%%%%%%%%%%%%%%%%%%%%%%%%
%%%%%%%%%%%%%%%%%%%%%%%%%%%%%%%%%%%%%%%%%%
%%%%%%%%%%%%%%%%%%%%%%%%%%%%%%%%%%%%%%%%%%

\subsection{Input-output fields calculations }

The exact
relation between input and output fields, Eqs. \eqref{rin} and
\eqref{rout}  is \cite{Gardiner1985} 
\begin{equation}
\label{io0}
r_ {\rm out} (t) = r_{\rm in} (t) - i 
\int_0^\infty  \frac {{\rm d} \omega }{\sqrt{2 \pi}}
\int_{t_0}^{t_f}  {\rm d} \tau \;  D (\omega) {\rm e}^{-i \omega (t-
  \tau)} X (\tau)
\; .
\end{equation} 
If we assume  that the coupling to the line is small compared to
the the scatterer transitions, only photons close to resonance (with
such  transitions) will actually interact with the scatterer.  
This is typical in experiments.   Then, we can approximate the
functional form $D(\omega)$ for its value at the incident photon  
frequency, $\omega$,  and \eqref{io0} is simplified:
\begin{equation}
\label{io}
r_ {\rm out} (t) = r_{\rm in} (t) - i  \sqrt{2 \pi } D (\omega) G (t)
\end{equation}
As a main consequence, $r_ {\rm out} (t)$ can be obtained 
by calculating the system dynamics.  It tuns out that, with the same
assumption yielding \eqref{io},  $G(t)$ can be obtained through the
quantum optical master equation \cite{Gardiner1985},
\begin{align}
\label{qme-0}
\frac {d \varrho}{dt}
= &
-i [H_{\rm sct}, \varrho]
-  i x(t)
[G , \varrho]
\\ \nonumber
&+
2  
\sum_{\omega_{ij} > 0}
\Gamma (\omega_{ij} )
\Big ( 
L_{ij} \varrho L_{ij}^\dagger 
-
\frac{1}{2}
\{  L_{ij}^\dagger   L_{ij} , \varrho \} 
\Big )  \, .
\end{align}
where 
$
L _{ij} = |j \rangle \langle i |
$ and
\begin{equation}
\label{G}
\Gamma (\omega_{ij}) = 2 \pi D^2 (\Omega) g_{ij}^2 + \gamma_{ij}
\end{equation}
Here, $\gamma_{ij}$ are the decays to another environments.  
The (classical) driving, due to the coherent input state, enters in
the second term of \eqref{qme-0}. 
The driving due to the coherent input state \eqref{alp-in} is taken into
account in the second term of \eqref{qme-0}  via 
\begin{equation}
x(t) = 
\langle X(t) \rangle_{line}
=
{\rm Tr} \big ( X (t) \varrho_{line} (t_0) \big ) 
\end{equation}
here,  $X(t) = {\rm e}^{i H_{\rm line} t } \,  X  \, {\rm e}^{-i H_{\rm
    line} t }$,  with $\varrho_{\rm line} (t_0)$ is the state of the
line (already with the input state). 
In the case of a coherent state as input \eqref{alp-in},
\begin{align}
\label{bt}
x(t)  = 
2 \alpha  D (\omega) \cos (\omega t) 
\end{align}
To solve for $t^{(1)}(\omega)$ [Cf. Eq. \eqref{t1-c}]
equation \eqref{qme-0}  must be solved in the limit 
of weak driving: $\alpha \to 0$.

\subsection{Linear Response theory (LRT)}
\label{app:LRT}

We review the Linear Response Theory (LRT), See. {\emph e.g.}
Ref. \onlinecite[Chap. 6]{Garcia-Palacios2007}.
We rewrite  \eqref{qme-0} as,
\begin{equation}
\label{L0L1}
\partial \varrho 
=
{\mathcal L_0}
\varrho
+
\lambda \, f(t) \,  {\mathcal L_1}
\varrho
\end{equation}
with,
\begin{align}
\label{L0cyclic}
\mathcal { L}_0  \varrho
= &
-i [H_{\rm S}, \varrho ]
\\ \nonumber
& +
2
\sum_\Omega
\Gamma (\Omega) 
\Big ( 
G(\Omega ) \varrho \;G^\dagger (\Omega)
-
\frac{1}{2}
\{ G^\dagger (\Omega) G (\Omega) , \varrho \} 
\Big )  \, .
%end{align}
%and 
%\begin{align}
\\
\label{L1cyclic}
\mathcal { L}_1  \varrho
= &   f (t) [G, \varrho]
\end{align}
and,
\begin{equation}
\lambda = \alpha  \sqrt {\left|\frac{d \omega}{d k}\right| } \;  D(\omega)
\qquad
\quad
f(t) = {\rm e}^{-i \omega t}.
\end{equation}
LRT solves the above evolution up to first order in $\lambda$:
\begin{equation}
\varrho =
\varrho_0 + \lambda \varrho_1
\end{equation}
with ${\mathcal L}_0 \varrho_0 = 0$,  \emph{i.e.} in absence of perturbation the system is
in equilibrium.
Replacing the above 
in \eqref{L0L1}  we get (up to first order)
\begin{equation}
\partial_t \varrho_1
=
{\mathcal L_0}
\varrho_1
+
f(t)  {\mathcal L_1}
\varrho_0
\, .
\end{equation}
The solution ($\varrho_1 (-\infty) = 0$) is 
\begin{equation}
\label{r1}
\varrho_1 = 
\int_{-\infty}^t
{\rm d} s 
\, {\rm e}^{ (t-s) {\mathcal L_0} }
\,
f (s) \, {\mathcal L_1} \varrho_0
\, .
\end{equation}

The solution \eqref{r1} is used to compute averages. In particular the
one for $G$:
\begin{equation}
\label{Adef}
\Delta G (t) :=
\langle G \rangle (t) 
-
\langle G \rangle_0
\qquad
\qquad 
\langle G \rangle_0 \equiv {\rm Tr} ( G \varrho_0 ) 
\, ,
\end{equation}
obtaining,
\begin{equation}
\label{DAgral}
\Delta G (t) 
= \lambda \int_{-\infty}^{\infty}
{\rm d}s R (t-s) f (s)
\end{equation}
with the \emph{response function} ($\theta (t) = 0$ if $t<0$, $\theta
(t) = 1$ otherwise)
\begin{equation}
R (t) = \theta (t)  \;  {\rm Tr}  ( \, {\rm e}^{ (t-s) {\mathcal L_0} }
\,  {\mathcal L_1} \varrho_0 ) 
\, .
\end{equation}
Importantly enough,  the response function $R (t)$ does not depend on
the actual form for $f(t)$.

\subsection{Practical calculation}

We discuss two important perturbation functions $f(t)$. We also
give the relation between them.

\subsubsection{Retarded perturbation}

In the, so called retarded perturbation, the 
perturbation $\mathcal L_1$ 
switched off at $t=0$.  Therefore, 
\begin{equation}
f_\text{r} (t) = \theta (-t)
\end{equation}
In this case \eqref{DAgral}
\begin{align}
\label{Ar}
\Delta G_\text{r} (t) &= 
 \lambda \int_{-\infty}^{\infty}
{\rm d}s R (t-s)
\theta (-s)
\\ \nonumber
&=
\lambda \int_{-\infty}^{\infty}
{\rm d}s 
R (s) \theta (s-t)
\\ \nonumber
&
=
\lambda
\int_t^\infty 
{\rm d}s R (s)
\quad  \longrightarrow \quad 
\frac{d}{dt} \Delta G_\text{r} = - \lambda R (t) 
\end{align}

\subsubsection{AC-driving}

We consider now, the  AC-driving $f(t) = {\rm e}^{-i \omega t}  $. 
%, starting at $t=0$,  $f(t) = {\rm e}^{i \omega t} \, \theta (t) $. 
 In this case, 
\begin{align}
\label{Aac}
\Delta G_\text{AC}(t) & = 
\lambda \int_{-\infty}^{\infty}
{\rm d}s R (t-s)
 {\rm
  e}^{-i \omega s}
\\ \nonumber
 & =
\lambda 
\,
{\rm
  e}^{-i \omega t}
\int_{-\infty}^\infty 
{\rm d} s
R (s) {\rm
  e}^{i \omega s}
\\ \nonumber
 &
%\stackrel {R(s<0) = 0} 
=
\lambda 
\,
{\rm
  e}^{-i \omega t}
\int_{0}^\infty 
{\rm d} s
R (s) {\rm
  e}^{ i \omega s}
\equiv
\lambda G (\omega )  {\rm
  e}^{-i \omega t}
\, .
\end{align}
The last equality defines the susceptibility.

\subsubsection{The relation}

Finally, both results the AC susceptibility and the retarded
evolution, Eqs. \eqref{Ar}  and \eqref{Aac} can be related,
\begin{align}
\label{AacAr}
\Delta G_\text{AC}(t)
& =
-  {\rm e}^{-i \omega t}
\Big ( 
\int_{0}^\infty 
{\rm d}  s
\, 
\frac{ 
d }
{ds}\Delta G_\text{r} 
\, 
 {\rm
  e}^{i \omega s}
\Big ) 
\\ \nonumber
 & =
- {\rm e}^{-i \omega t}
\Big ( 
\int_{0}^\infty 
{\rm d}  s
\, 
\frac{ 
d G_\text{r} }
{ds}
\, 
 {\rm
  e}^{ i \omega s}
\Big )
\\ \nonumber
&=
 {\rm e}^{-i \omega t}
\Big ( 
G_\text{r} (0) 
+
i \omega  
\int_{0}^\infty 
{\rm d}  s
\, 
 G_\text{r} (s) 
\, 
 {\rm
  e}^{ i \omega s}
\Big ) 
\end{align}
Here $G_\text{r}(\infty) = 0$.

Summarazing, for computing the evolution under AC driving we do not
need to solve the explicit time dependent problem \eqref{L0L1} but
solve the unperturbed evolution with initial conditions $ ( \mathcal {
L}_0  + \lambda \mathcal {L}_1 ) \varrho = 0$ (the retarded response).

\subsection{Final formula}

The solution for $\partial_t \varrho = {\mathcal L}_0
\varrho$ is
\begin{equation}
\label{Ars}
G_\text{r} (s) = \sum_{ij} G_{i,j} \varrho_{ij} (0) {\rm e}^{-(i \omega_{ij} +
  \Gamma_{ij} ) s}
\end{equation}
with $G_{i,j} = \langle i | G | j \rangle$  and $\varrho_{ij} (0) $ the initial conditions obtained by solving
$ ( \mathcal {
L}_0  + \lambda \mathcal {L}_1 ) \varrho = 0$.
Eq. \eqref{AacAr} can be rewritten (and approximated) in energy space
\begin{align}
\nonumber
\Delta G_\text{AC}(\omega)  &=
\sum_{ij} G_{i,j} \varrho_{ij} (0)
\left (1 + \frac{ i \omega}{ i (\omega_{ij} - \omega)  + \Gamma_{ij}} 
\right )
\\ \nonumber
&=
 \sum_{ij} G_{i,j} \varrho_{ij} (0)
 \frac{ i \omega_{ij} + \Gamma(\omega_{ij})}{ i (\omega_{ij} - \omega) + \Gamma_{ij}} 
\\  \label{AacRWA}
& \cong
\sum_{\omega_{ij} >0} G_{i,j} \varrho_{ij} (0)
\frac{ i \omega_{ij}+  \Gamma(\omega_{ij})}{ i (\omega_{ij} - \omega) + \Gamma_{ij}} 
\end{align}
The last approximation considers that the main contribution comes from
the terms with  poles.

\begin{widetext}
%%%%%%%%%%%%%%%%%%%%%%%%%%%%%%%%%%
%%%%%%%%%%% T1     %%%%%%%%%%%%
%%%%%%%%%%%%%%%%%%%%%%%%%%%%%%%%%%
\subsection{Transmission calculation}
\label{app:T1}

Following   Eq. \eqref{AacRWA}, we can approximate
\begin{equation}
G_\text{r} (s) \cong  
G_{0 1 } \varrho_{1 0 } (0) {\rm e}^{-(i \omega_{01}+
  \Gamma (\omega_{01}) ) s}
+
G_{0 2 } \varrho_{2 0 } (0) {\rm e}^{-(i \omega_{02}+
  \Gamma (\omega_{02}) ) s}
\end{equation}

Finally, we solve for   $\varrho_{ij} (0)$, that are solutions of 
$ ( \mathcal {
L}_0  + \lambda \mathcal {L}_1 ) \varrho = 0$.
We get the equations,
%\begin{widetext}
%\begin{align}
%\label{dr10}
%\dot \varrho_{10} = 0
%&= 
%- i \Big  ( \omega_{01} + \alpha(S_{11} - S_{00} )
%\Big ) \varrho_{10}
%- i \alpha S_{10} ( \varrho_{00} -  \varrho_{11} ) 
% - 
%\Gamma_{01} 
% \varrho_{20}
%\\ \label{dr20}
%\dot \varrho_{20} = 0 
%&= - i \Big ( \omega_{02} + b (S_{22} - S_{00} ) \Big ) \varrho_{20}
%- i b S_{20} ( \varrho_{00} -  \varrho_{22} ) 
% - 
%\Big ( 
%\Gamma_{02}
%+
% \Gamma_{01}
% \Big ) \varrho_{20}
%\end{align}
\begin{align}
\label{dr10}
\dot \varrho_{10} = 0
&= 
- i \omega_{01}  \varrho_{10}
- i \alpha g_{10} ( \varrho_{00} -  \varrho_{11} ) 
 - 
\Gamma_{01} 
 \varrho_{20}
\\ \label{dr20}
\dot \varrho_{20} = 0 
&= - i  \omega_{02} \varrho_{20}
- i x(t) g_{20} ( \varrho_{00} -  \varrho_{22} ) 
 - 
\Big ( 
\Gamma_{02}
+
 \Gamma_{01}
 \Big ) \varrho_{20}
\end{align}
and the ones for the diagonals
\begin{align}
\label{00}
\dot \varrho_{00} =  0  & =
-i \alpha g_{10} ( \varrho_{01} -\varrho_{10} )
-i \alpha g_{20} ( \varrho_{02} -\varrho_{20} )
+ \Gamma_{01} \rho_{11}  + \Gamma_{02} \rho_{2 2}
\\
\label{11}
\dot \varrho_{11} =  0  & =
+i \alpha g_{10} ( \varrho_{01} -\varrho_{10} )
-i \alpha g_{21} ( \varrho_{12} -\varrho_{21} )
- \Gamma_{01} \rho_{11}  + \Gamma_{12} \rho_{2 2}
\end{align}
and $\varrho_{22} = 1 - \varrho_{11} - \varrho_{00}$.
%\end{widetext}
From \eqref{00} and \eqref{11} we see that,
$\varrho_{11} \sim \varrho_{22} \sim {\mathcal O} (\alpha)$ and
$\varrho_{00} \sim 1- {\mathcal O} (\alpha)$ 
wich makes trivial solve \eqref{dr10} and \eqref{dr20} up to first
order in $\alpha$.
Inserting their solutions in the general
expression \eqref{AacRWA} we get the equation in main text.

\end{widetext}

%%%%%%%%%%%%%%%%%%%%%%%%%%%%%%%%%%
%%%%%%%%%%% T1     %%%%%%%%%%%%
%%%%%%%%%%%%%%%%%%%%%%%%%%%%%%%%%%
\subsection{Leakage}
\label{app:tau}

%%%%%%%%%%%%%%%%%%%%%%%%%%%%%%%%%%%
\begin{figure}
\includegraphics[width=0.7\columnwidth]{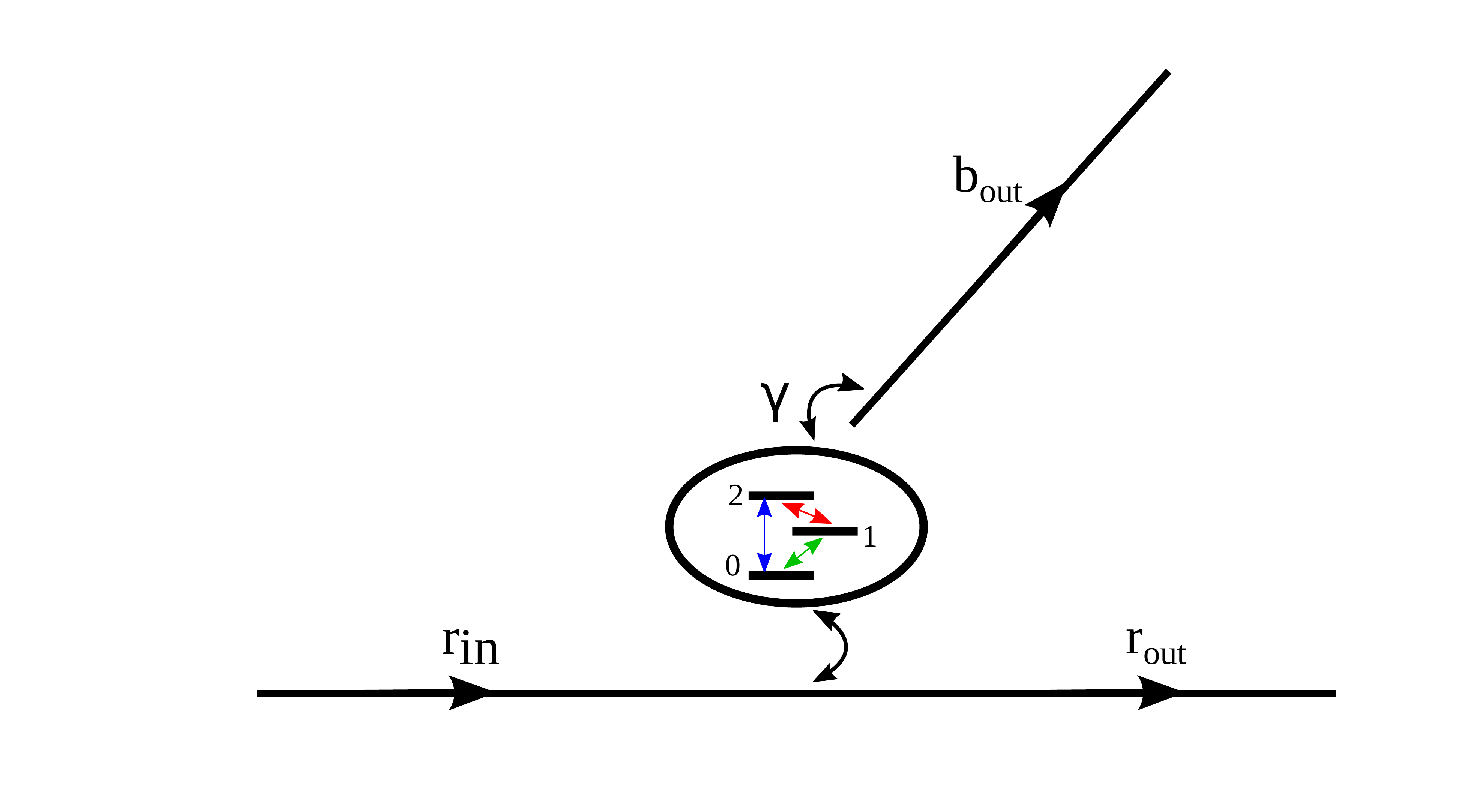}
 \caption{(Color online) Schematics for the modeling of non-radiative losses}
 \label{fig:losses}
 \end{figure}
%%%%%%%%%%%%%%%%%%%%%%%%%%%%%%%%%%%
Losses can be modelled as decays to another channels.  Here, we 
take into account one channel (others will sum up), see
Fig. \ref{fig:losses}.  
The input-ouput relations \eqref{io} must be generalized now to
include this \emph{extra} channel,
\begin{align}
r_ {\rm out} (t) = & r_{\rm in} (t) - i  \sqrt{2 \pi } D (\omega) G (t)
\\
b_{\rm out} (t) = & b_{\rm in} (t) - i  \sqrt{ 2 \gamma} G (t) \, .
\end{align}
The $ \gamma$ gives a phenomenological loss rate, and the $2$ in front
is because we do not consider left and right modes in the non radiative
channel but just $b$-modes.
Besides, $b_{\rm in} (t) =0$ and the transmission in the
$b$-modes read,
\begin{equation}
\tau (\omega)  = 
 \frac{-i \sqrt{2 \gamma} \langle G \rangle}{\langle
  r_{\rm in} \rangle}  
\end{equation}

\section{Efficiency calculations}
\label{app:eff}

\begin{figure*}[t]
\includegraphics[width=1.8\columnwidth]{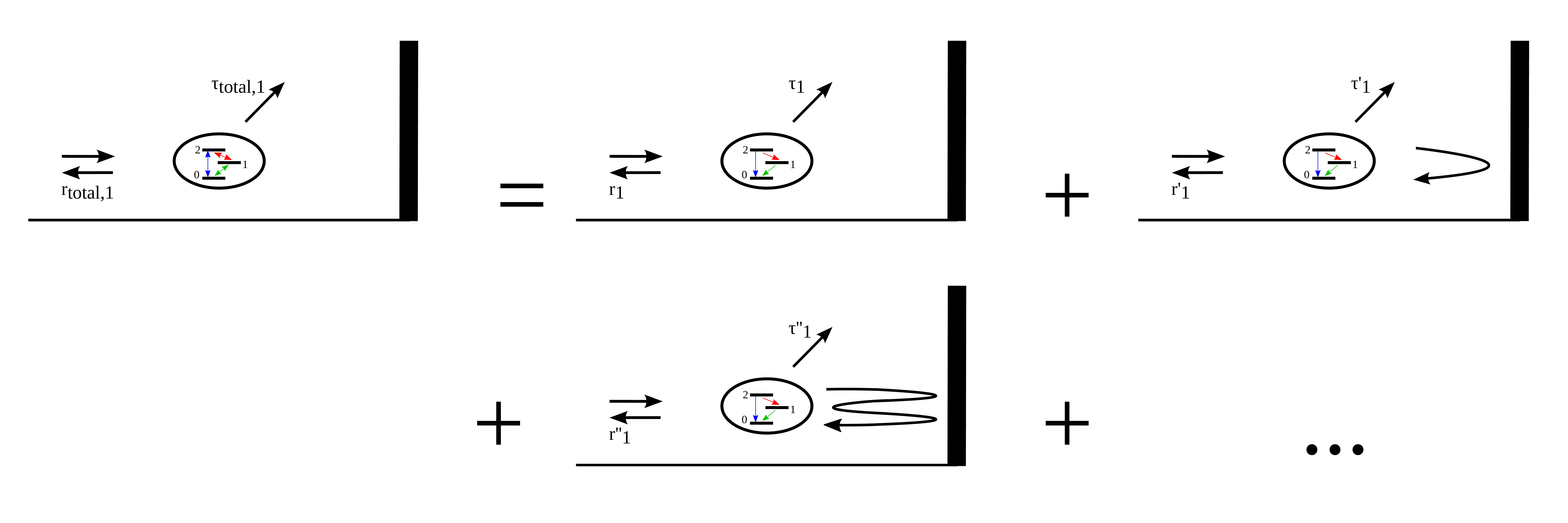}
 \caption{(Color online) Diagrammatic plot for the possible scattering
 events giving the total reflection}
 \label{fig:events}
 \end{figure*}

In order to compute the reflection and leakage when the mirror is
placed,  we must sum over all the possible reflection, transmission and leakage
events, as shown in Figure \ref{fig:events}.
In doing so, we  name $\Phi (\omega)= {\rm exp} ( i k (\omega) d) $ the phase accumulated by a photon
with quasi momentum $k$
travelling a distance $d$ (this will be the distance between the
mirror and the atom). 
Finally, we denote the reflection in the mirror as $r_M$.  Eventually,
we will set $r_M=-1$, \emph{i.e.} we neglect losses in the mirror, wich is a good
experimental assumption. 

%%%%%%%%%%%%%%%%%%%%%%%%%%%%%%%%%%%
%%%%%%%%%%%%%%%%%%%%%%%%%%%%%%%%%%%

With the mirror, $P_2 (\omega)$ is written as,
\begin{equation}
\label{P2l}
P_2 (\omega) = 1 - |r_{\rm tot,1} (\omega)|^2 -  |\tau_{\rm tot} (\omega)|^2
\end{equation}
where $r_{\rm tot,1} (\omega)$ is the \emph{total} one photon reflection.  It should be 
 distinguished  from $r_1$ wich stands for the reflection occurring in every
  event.  Finally, $\tau_{\rm tot} (\omega) $ is the total leakage.
Summing over all the events, see Figure \ref{fig:events}, we finally get
\begin{align}
\nonumber
r_{{\rm tot}, 1} (\omega)  = & r_1 (\omega) + t^{(1)}(\omega) \Phi(\omega) r_M t^{(1)}(\omega)
\\ \nonumber
& +
 t^{(1)}(\omega) \Phi(\omega) r_M  t^{(1)}(\omega)r_1 (\omega) \Phi(\omega) r_M t^{(1)}(\omega)
+ 
\ldots
\\  \label{r1tot}
= &
r_1 (\omega) + \frac{ (t^{(1)}(\omega))^2 \Phi(\omega) r_M }{1 - r_1 (\omega) \Phi(\omega) r_M },
\end{align}
and 
\begin{align}
\nonumber
\tau_{\rm tot} (\omega)
= &
\tau (\omega)
+ t^{(1)}(\omega) \Phi(\omega) r_M \tau (\omega)
\\ \nonumber
& +
 t^{(1)}(\omega) \Phi(\omega) r_M t^{(1)}(\omega)r_1 (\omega)
  \Phi(\omega) r_M  \tau (\omega)
+ 
\ldots
\\  \label{tautot}
= &
\tau (\omega) + \frac{ \tau (\omega) t^{(1)}(\omega) \Phi(\omega) r_M }{1 - r_1
    (\omega) \Phi(\omega) r_M}
\, .
\end{align}
Combining \eqref{r1tot}, \eqref{tautot} with \eqref{P2l} we can
compute the two photon generation $P^{(2)} (\omega)$, considering
$r_M=-1$. In the main text, we introduce 
\begin{equation}
A(\omega) \equiv |\tau(\omega)|^2 \, .
\end{equation}

\section{Numerical simulations}
\label{sec:MPS}
\subsection{Matrix Product States}

We are studying the dynamics of a state with
one or two photons flying over the ground state, \emph{i.e.} the state
is expected to have a small amount of entanglement.  Therefore, 
 we can use the variational ansatz of Matrix Product States
 \cite{JJRipoll2006,Verstraete2008} to describe the discrete wave
 function as we have shown recently \cite{Burillo2014, Sanchez-Burillo2015}.  This ansatz has the form
\begin{equation}
\ket{\psi} = \sum_{s_x \in \{1,d_x\}} \mathrm{tr}\left[
\prod A_x^{s_x}\right] \ket{s_1,s_2,\ldots,s_L}.
\label{eq:mps}
\end{equation}
It is constructed from $L$ sets of complex matrices $A_x^{s_x} \in M[\mathbb{C}^{D}]$, with $L$ the number of sites, where each set is labelled by the quantum state $s_x$ of the corresponding site. The local Hilbert space dimension $d_x$ is infinity, since we are dealing with bosonic sites. During the dynamics, processes that create more than two photons are still highly off-resonance. In consequence, we can truncate the bosonic space and consider states with $0$ to
$n_{max}$ photons per cavity. The composite Hilbert space is $\mathcal{H}=\bigotimes_x \mathbb{C}^{d_x}$, where the dimension is $d_x=n_{max}+1$ for the empty resonators and $d_{x_0}=3(n_{max}+1)$ for the cavity with the three-level system. We thus expect the composite wave function of the photon-C3LS to consist of a superposition with a small number of photons

The total number of variational parameters $(L-1)D^2(n_{max}+1) + 3D^2(n_{max}+1)$ depends on the size of the matrices, $D$. The key point is that, for describing a general state, $D$ increases exponentially with $L$, whereas it increase polynomially if the entanglement is small enough.

Our work with MPS uses four different algorithms. The most basic one
is to create product states such as a vacuum state with the deexcited
C3LS: $\ket{\psi}=\ket{0}\ket{\text{vac}}$. These states can be
reproduced using matrices of bond dimension $D=1$, so each matrix is
just a coefficient $A_x^{s_x}=\delta_{s_x1}$. The second algorithm is
to compute expectation values from MPS. This amounts to a contraction
of tensors that can be performed efficiently \cite{JJRipoll2006}, and
allows us to compute single-site operators $\langle a^\dagger_x
a_x\rangle$, $\langle \sigma_z\rangle$, correlators as $\langle
a_x^\dagger a_x\rangle$ or even projections as $\langle\Omega
|a_{x_1}a_{x_2}|\psi\rangle$. The third operation that we need to
perform is to apply operators on to the state, $O\ket{\psi}$, such as
introducing or removing excitations $a_x^\dagger\ket{\psi}$. We do
this in an efficient fashion by interpreting the operator $O$ as a
Matrix Product Operator (MPO) \cite{Verstraete2010}. A MPO is a matrix
product representation of an operator:
\begin{widetext}
\begin{equation}
O = \sum_{s_{x}^{},s_{x}^\prime \in \{1,d_x\}} \mathrm{tr}\left[\prod B_x^{s_x^{},s_x^\prime}\right]\ket{s_1^{},s_2^{},\ldots,s_L^{}}\bra{s_1^\prime,s_2^\prime,\ldots,s_L^\prime}
\end{equation}
\end{widetext}
So, now we have $L$ sets of complex matrices $B_x^{s_x^{},s_x^\prime} \in M[\mathbb{C}^{D_O}]$, where each set is labelled by two indices $s_x^{},s_x^\prime$ of the corresponding site.

We just need to apply sums of one-body operators

\begin{equation}\label{eq:one_body_op}
O = a_\phi^\dagger = \sum_x \phi_x a_x^\dagger.
\end{equation}

In such a case, an efficient representation of the MPO is obtained with $D_O=2$

\begin{equation}
B_x^{s_x^{},s_x^\prime}=\left(\begin{array}{c c}
\delta_{s_x^{},s_x^\prime} & 0\\
\phi_x(a_x^\dagger)_{s_x^{},s_x^\prime} & \delta_{s_x^{},s_x^\prime}
\end{array}\right)\qquad x=2,3,\dots,L-1,
\end{equation}
whereas $B_1^{s_1^{},s_1^\prime}=(\phi_1(a_1^\dagger)_{s_1^{},s_1^\prime},\delta_{s_1^{},s_1^\prime})$ and $B_L^{s_L^{},s_L^\prime}=(\delta_{s_L^{},s_L^\prime},\phi_L(a_L^\dagger)_{s_L^{},s_L^\prime})^T$, with $(a_x^\dagger)_{s_x^{},s_x^\prime}=:\bra{s_x^{}}a_x^\dagger\ket{s_x^\prime}$.

Finally, we can also approximate time evolution, repeatedly contracting the state with an MPO approximation of the unitary operator $\exp(-iH\Delta t)$ for short times, and truncating it to an ansatz with a fixed $D$. Since our problem does not contain long-range interactions and since the state is well approximated by MPS, it is sufficient to rely on a third-order Suzuki-Trotter formula \cite{Suzuki1991}.
In the same way as we can consider time evolution, we can take imaginary time to obtain the ground state, that is solving the equation $\frac{d}{dt}\ket{\psi}=-H\ket{\psi}$ for finite time-steps, while constantly normalizing the state. Provided a suitable initial state, the algorithm converges to the lowest-energy state of $H$. Notice that the ground state is totally necessary to study the dynamics, since our initial state is obtained by applying a single-body operator as that of Eq. \ref{eq:one_body_op} over the ground state.

\subsection{Simulated model, input state and parameters used}

For the photonic medium, we consider a one-dimensional array of coupled cavities:
\begin{align}
H= & \epsilon \sum_x a_x^\dagger a_x - J\sum_x (a_x^\dagger a_{x+1} +
     \text{H.c.})
\\ \nonumber
&+\sum_i \omega_i |i\rangle\langle i| + \sum_{ij}(g_{ij}\ket{i}\bra{j}+\text{H.c.})(a_0+a_0^\dagger),
\end{align}
being $\epsilon$ the bare frequencies of the cavities, $J$ the hopping between nearest neighbours and $g_{ij}$ the coupling constant for the $\ket{i}\leftrightarrow\ket{j}$ transition. The lattice spacing $d$ is fixed to 1. The photonic part can be diagonalized in momentum space, giving the dispersion relation $\omega(k)=\epsilon-2J\cos k$. The density of electromagnetic modes will be $D(\omega)=1/\sqrt{2J|\sin (k(\omega))|}$.

We fix $\epsilon=1$, $J=1/\pi$, $\omega_0=0$, $\omega_1=0.59$ and $\omega_2=1.10$ (these energies were obtained from the model introduced in the main part of the text). We take $L=1000$ cavities and we place the scatterer at $x_0=500$ (in the main text, we consider $x_0=0$). The couplings used in the simulations to compute the full spectrum are $g_{01}=-0.0225$, and $g_{12}=g_{02}=0.03$, which were obtained from the physical implementation we shall explain below. In the simulations in which we computed the two-photon wave function, in order to get a cleaner scattering state and due to limitations in the time of simulation, we artificially increased the couplings: $g_{01}=-0.10$, $g_{12}=g_{02}=0.13$.

We work in position space. The input state is:
\begin{equation}
|\Psi_\text{in}\rangle=\sum_x e^{ik_0 x}e^{(x-\bar{x})/2\sigma}\theta(x_0-x) a_x^\dagger|\Omega\rangle,
\end{equation}
up to a normalization constant, with $\bar{x}$ the position of the wave front, $\sigma$ the width, $k_0$ the mean momentum and $\theta(x)$ the Heaviside function. We fix $\bar{x}=420$ and $k_0=1.73$ (on resonance with $\omega_{02}$). We take $\sigma=2$ for the simulations to get the full spectrum and $\sigma=20$ for the simulation in which we compute the two-photon wave function.
The results reported used bond dimension  $D=10$ and the cut-off for the
cavities is $n_\text{max}=3$.  We checked that these sizes are already sufficient.

\end{appendix}

%%%%%%%%%%%%%%%%%%%%%%%%%%%%%%%%%%
%%%%%%%%%%% BIBLIOGRAPHY    %%%%%%%%%%%%
%%%%%%%%%%%%%%%%%%%%%%%%%%%%%%%%%%

\bibliographystyle{apsrev4-1}
\bibliography{scattering_eduardo}

%%%%%%%%%%%%%%%%%%%%%%%%%%%%%%%%%%
%%%%%%%%%%% the end     %%%%%%%%%%%%
%%%%%%%%%%%%%%%%%%%%%%%%%%%%%%%%%%

\end{document}